\documentclass[%
nofootinbib,
 amsmath,amssymb,
 aps,
 prd
]{revtex4-1} 

\usepackage[utf8]{inputenc}

\usepackage{amsmath}
\usepackage{dcolumn}
\usepackage{bm}
\usepackage{csquotes}
\usepackage{graphicx}
\usepackage[english]{babel}
\usepackage[T1]{fontenc}
\usepackage{amssymb}
\usepackage{setspace}
\usepackage{lmodern}
\usepackage{geometry}
\usepackage{enumerate}
\usepackage{caption}
\usepackage{subcaption}
\usepackage{algorithm}
\usepackage[]{algpseudocode}
\usepackage[dvipsnames]{xcolor}
\usepackage[normalem]{ulem}
\usepackage{hyperref}
\hypersetup{
    colorlinks,
    citecolor=blue,
    filecolor=blue,
    linkcolor=blue,
    urlcolor=blue
} 
\usepackage{soul}

\newcommand{\af}[1]{\textcolor{black}{#1}} 
\newcommand{\vm}[1]{\textcolor{black}{#1}} 
\newcommand{\om}[1]{\textcolor{black}{#1}} 

\usepackage{natbib}
\begin{document}


\title{Testing the mass of the graviton with Bayesian planetary numerical ephemerides B-INPOP}

\author{Vincenzo \surname{Mariani}}
 \email{vmariani@geoazur.unice.fr}
\affiliation{%
Universit\'e C\^ote d'Azur, Observatoire de la C\^ote d'Azur, CNRS, 250 av. A. Einstein, 06250 Valbonne, France}
\author{Agn\`es  \surname{Fienga}}%
 \email{agnes.fienga@oca.eu}
\affiliation{%
Universit\'e C\^ote d'Azur, Observatoire de la C\^ote d'Azur, CNRS, 250 av. A. Einstein, 06250 Valbonne, France,
IMCCE, Observatoire de Paris, PSL University, CNRS, 77 av. Denfert-Rochereau, 75014 Paris, France}%


\author{Olivier \surname{Minazzoli}}
 \email{ominazzoli@gmail.com}
\affiliation{
Artemis, Universit\'e Cote d’Azur, CNRS, Observatoire Cote d’Azur, BP4229, 06304, Nice Cedex 4, France,
Bureau des Affaires Spatiales, 2 rue du Gabian, 98000 Monaco
}%
%

\begin{abstract}
We use MCMC to sample the posterior distribution of the mass of the graviton ---assumed here to be manifest through a Yukawa suppression of the Newtonian potential---by using INPOP planetary ephemerides. The main technical difficulty is  the lack of analytical formulation for the forward problem and the cost in term of computation time for its numerical estimation.
To overcome these problems we approximate an interpolated likelihood for the MCMC with the Gaussian Process Regression. We also propose a possible way to assess the uncertainty of approximation of the likelihood by mean of some realization of the Gaussian Process. 
At the end of the procedure, a 99.7\% confidence level threshold value is found at $1.01 \times 10^{-24} \; eV c^{-2}$ (resp. $\lambda_g \geq 122.48 \times 10^{13} \; km$), representing an improvement of 1 order of magnitude relative to the previous estimation of Bernus et al. \cite{Bernus2020}. Beyond this limit, no clear information is provided by the current state of the planetary ephemerides.
\end{abstract}

\maketitle

\date{\today}

\section{Introduction}

Planetary ephemerides have evolved with the astrometric accuracy obtained for the astrometry of planets and natural satellites thanks to the navigation tracking of spacecraft (s/c) orbiting these systems. Since the late XIXth century the astrometry of planets has known a significant improvement leading to an increased accuracy of the dynamical theories describing their motions. 
The motion of the planets and asteroids in our Solar System can be solved directly by the numerical integration of their equations of motion.
The improved (present and future) accuracy in the measurements of observables from space missions like Cassini–Huygens, MEX, VEX, BepiColombo etc. makes the Solar System a suitable arena to test General relativity theory (GRT) as well as alternative theories of gravity by mean of Solar System ephemerides.  
INPOP planetary ephemerides are developed since 2003, integrating numerically the Einstein-Infeld-Hoffmann equations of motion proposed by \cite{moyer1971mathematical, moyer2003monography},  and fitting the parameters of the dynamical model to the most accurate planetary observations following \cite{Soffel_2003}.
Testing alternative theories of gravity with planetary ephemerides  consist in changing the metric of GRT into alternative frameworks and consequently modifying the equations of
motion, the light time computation and the definition of time-scales used for the construction of planetary ephemerides.  In principle, such modifications of GRT can be {summarized as} considering additional terms to GRT fundamental equations,
such as {a Yukawa suppression of the Newtonian potential}.

{At the phenomenological level, a mass of the gravitational interaction\footnote{{Reported as \textit{the mass of the graviton} in the rest of the paper, for convenience.}} is often assumed to either lead to modification of the dispersion relation of gravitational-waves \cite{LIGOVirgo_testsGRT_12_2021} or to lead to a Yukawa suppression of the Newtonian potential \cite{PhysRevD.57.2061}. \footnote{{For more information on the status of current theoretical models of massive gravity, we refer the reader to \cite{RevModPhys.89.025004, DeRham_MassiveGravity}}.}}


{Recently \cite{Will_2018},\footnote{{Following his more-than-twenty years old seminal work \cite{PhysRevD.57.2061}.}}}
Will argued that Solar System observations and planetary dynamics could be used to improve the constraints on the mass of the graviton $m_g$, obtained from the LIGO-Virgo Collaboration. However Will uses results based on statistics of \om{postfit} residuals of the Solar System ephemerides that are performed without including the effect of a massive graviton inside the equations of motion. In order to overcome \om{the consistency issues that are raised by such type of analyses---that is, which are based on postfit residuals---}
we investigate a new approach \om{that is based on} a statistical inference \om{of} the mass of the graviton $m_g$ within the framework of INPOP as presented in Bernus et al. \cite{bernus2019, Bernus2020}.

The approach we decided to use is partially Bayesian and several tools like MCMC, Gaussian Process and Bayes Factor are employed, in order to exploit at the best the values of INPOP $\chi^2$. This work is a generalization of what had already be started by \cite{Bernus2020} with some differences and improvements.

In this work we use the INPOP ephemerides INPOP21a \cite{INPOP21a}. We first introduce the method and how we overcome the problem of the time cost for the forward model. 
{In Sec. \ref{sec:Methods}, we describe the specific algorithms we used. We give in Sec. \ref{sec:Results} the result after a full convergence of the planetary ephemerides and we give }\af{ a new limit at 99.7$\%$ confidence level (CL) for the mass of the graviton. For masses smaller than this threshold, we show no clear sensitivity of  the plenetary ephemerides.} 
 Finally, in Sec. \ref{sec:Discussion}, we compare our new results in particular considering former publications from Bernus. 

\section{Methods}\label{sec:Methods}

\subsection{Planetary ephemerides construction}
\label{sec:inpop}
INPOP (Int\'egrateur Num\'erique Plan\'etaire de l’Observatoire de Paris) is a planetary ephemerides that is built by integrating numerically the equations of motion of the Solar System objects following the formulation of Moyer \cite{moyer2003monography}, and by adjusting to Solar System observations such as {space mission navigation and radio science data, ground-based optical observations or lunar laser ranging} (\cite{Fienga2008, INPOP21a}).
In addition to adjusting the astronomical intrinsic parameters, it can be used to constrain parameters that encode deviations from GRT \cite{inpop10a_ff, verma2014, Fienga-2015, Viswanathan2018}, such as the Compton wavelength $\lambda_g$. This is defined such as 

\begin{equation} \label{eq:def_mg_lambda}
\lambda_g = \frac{\hbar}{c m_g}
\end{equation}
with $\hbar$ the Planck constant.
As long as $m_g$ is small enough, the gravitational phenomenology in the Newtonian regime {recovers the} one of GRT. 
{Recently} \cite{Will_2018}, Will argued that Solar System observations could be used to improve---or at least be comparable with---the constraints on $m_g$ obtained from the LIGO-Virgo Collaboration---assuming that the parameters $m_g$ appearing in both the radiative and Newtonian limits are the same. 
A graviton mass would indeed lead to a modification of the perihelion advance of Solar System bodies. 
Based on current constraints on the perihelion advance of Mars---or on the post-Newtonian parameters $\gamma$ and $\beta$---derived from Mars Reconnaissance Orbiter (MRO) data, Will estimates that the graviton  mass should be smaller or equal to  $(5.6 - 10) \times 10^{-24} \; eVc^{-2}$ depending on the specific analysis. 
However, as an input for his analysis, Will uses results based on \vm{interpreting} statistics of \vm{postfit} residuals of the Solar System ephemerides \vm{obtained in \om{various frameworks (PPN and GRT)} as possible outcome of the graviton influence. 
\om{However, first of all---unlike the historical occurrence of the substantial error in the perihelion advance of Mercury computed in Newton's theory---}a lot of different contributions \om{from the details of the Solar System model being used} could explain the \om{rather small} postfit differences between computed and observed positions.}
Furthermore various parameters of the ephemerides (e.g., masses, semimajor axes, etc.) are all more or less correlated to $m_g$ as it is shown in \cite{bernus2019}. 
Therefore any kind of signal introduced by $m_g > 0$ can, in part, be reabsorbed during the fit of other correlated parameters. In order to \af{overcome \om{the} correlation issues decribed previously,} we investigate a new approach based on a statistical inference on the mass of the graviton $m_g$ within the full framework of INPOP. 

Considering the $N-$body system, the acceleration included in the INPOP \vm{modified} code is given by \cite{bernus2019}
\begin{equation} \label{eq:accelaration_graviton}
\delta a^i = \frac{1}{2} \sum_{P} \frac{ c^2 GM_P m_g^2 }{h^2} \frac{x^i - x^i_P}{r} + \mathcal{O}(m_g^{3}). 
\end{equation}
\vm{In Eq. \eqref{eq:accelaration_graviton}}, $M_P$ and $x^i_P$ are the mass and the position of the gravitational source $P$, \vm{whereas $x^i$ is the position of the body subject to the gravitational force of $P$ and $r$ is the distance between the two objects}. This model has been already implemented and is described in Bernus et al. \cite{bernus2019, Bernus2020}.
In this work we will use the INPOP21a planetary ephemerides \cite{INPOP21a}. This version benefits from the latest Juno and Mars orbiter tracking data up to 2020 as well as a fit of the Moon-Earth system to LLR observations also up to 2020. For a more detailed review about this specific version, the reader can refer to \cite{INPOP21a} whereas \cite{2022IAUS..364...31F} gives more descriptions regarding recent GRT tests obtained with INPOP21a and INPOP19a. INPOP21a is more accurate than INPOP19a especially for Jupiter and Saturn orbits as additional Juno observations of Jupiter were used covering a 4-years period when only 2.5 years were considered in INPOP19a. Consequently, a more realistic model of the Kuiper belt was implemented in INPOP21a, leading to an improvement of about 1.4 on the Jupiter orbit accuracy. Additionally, 2 years of Mars Express navigation data have been added to the 13 years already implemented in INPOP19a. This increase of Mars orbiter data improves mainly the stability of the ephemerides and its extrapolation capabilities \cite{10.1093/mnras/stz3407}. In terms of adjustment, in addition to the initial conditions of the planetary orbit,  the gravitational mass of the Sun, its oblateness and the ratio between the mass of the Earth and the one of the Moon, 343 asteroid masses are fitted in INPOP21a following the procedure described, for example, in \cite{10.1093/mnras/stz3407}. A mass representing the average effect of 500 trans-neptunian objects has also been added as described in \cite{INPOP21a}. A total of 401 parameters are accounted for the INPOP21a construction. They constitute the list of astronomical parameters we will refer to in the following.


\subsection{MCMC and Metropolis algorithms}
\label{sec:th}
In the past years there have been already some attempts to deal with the problem of high correlations among parameters inside the INPOP planetary ephemerides fit,
as in the case of the determination of asteroid masses (see \cite{10.1093/mnras/stz3407}). For testing alternative theories and thus assessing threshold values for the violation of GRT, \cite{Fienga-2015} had tested genetic algorithm approaches for identifying intervals of values for parameters such as PPN, $\beta$, $\gamma$, the Sun oblateness $J_2$ and secular variations of the gravitationnel mass of the Sun $\frac{\Dot{\mu}}{\mu}$, with which planetary ephemerides can be computed and fitted to the observations with a comparable accuracy than the ephemerides built in GRT.

Keeping in mind the problem of correlation between planetary ephemerides and GRT parameters, we propose a new procedure with a semi-bayesian approach to test a possible deviation from GRT in a particular case: we investigate the posterior probability distribution of a possible non-zero mass of the graviton $m_g$ employing MCMC techniques.

Our procedure is semi-bayesian in the sense that only the mass of the graviton is actually sampled with the MCMC procedure, \vm{the INPOP astronomical parameters being fitted with a least square procedure}. We follow here the algorithm already used by \cite{bernus2019,Bernus2020,2022IAUS..364...31F}: for a fixed value of $m_g$, we integrate the motion of the planets with INPOP and we fit to planetary observations, the astronomical parameters listed in Sec. \ref{sec:inpop} in using the least square iterative procedure described in \cite{10.1093/mnras/stz3407}. We then obtain a fully fitted ephemerides built for a fixed value of $m_g$. 
%
%

\subsubsection{The classic MH and MCMC}

Generally speaking, the Metropolis-Hastings (MH) algorithm is one of the first Markov Chain Monte Carlo methods developed, providing a sequence of random samples drawing them from a given probability distribution. In a first step, we are going to describe the procedure (the algorithm) in a generic framework (i.e. sampling a generic probability distribution) and then giving in detail which distribution we are sampling from. For a detailed overview on the MCMC method see \cite{brooks2011handbook} and \cite{RobertCasella2004}. 
We will not provide a proof of convergence of the method since it is out of the goal of our work and it can be found easily in \cite{RobertCasella2004} and  \cite{brooks2011handbook}.
The MH algorithm associated with the objective (target) density $f$ and the conditional density $q$ produces a Markov chain $(X_n)$ through the following transition.
Given the $n$-th element of the chain $X_n$, we first generate $Y_t \sim q(y|X_n)$. We then take $X_{n+1}$ with
  \[ X_{n+1} = \Biggl\{ \begin{array}{lcl} Y_t & \text{with probability} & p_a(X_n, Y_t) \\ X_n & \text{with probability} & 1 - p_a(X_n, Y_t) \end{array}   \] where 
    
    \begin{equation}\label{AccRatio_form1}
p_a(x,y) = \min \left\{ 1, \frac{f(y)}{f(x)} \frac{q(x | y)}{q(y|x)} \right\}
\end{equation}

    

 Iterating this drawing process for enough iterations we obtain a distribution that samples the target density $f$. These two operations represent an \emph{step} of the MCMC. All the steps together, with their output, yield to the Markov Chain we are seeking for. 

\subsubsection{MH algorithm in our case} \label{sec:MH_in_our_case}

The main idea is to use as target probability the posterior probability function of our inverse problem with respect to the observational data used in INPOP. Therefore the parameter for which we want the posterior is the mass of the graviton and in particular the probability distribution we are going to sample is:
\begin{equation}\label{eq:posterior_prior_likelihood}
\pi(m_g) \propto \rho(m_g) \times L(m_g)
\end{equation}
In Eq. \eqref{eq:posterior_prior_likelihood} $\rho$ is the prior density function for $m_g$ and $L$ is the likelihood. \vm{A more precise version of Eq. \eqref{eq:posterior_prior_likelihood} is given later on. }
From now on we will refer to the mass of the graviton with $m_g  \geq 0$ by assumption. 
Let's note that we \emph{just} need to know the posterior $\pi$ \emph{up to a constant} in the sense that the posterior in the algorithm will be used in Eq. \eqref{AccRatio_form1} to compute the MH acceptance probability. It then appears in a ratio, and knowing the posterior up to a constant value is enough.
We will use a uniform prior, i.e. a density function that has a constant non-zero value on his support\footnote{Here by support we mean the mathematical sense:  the support of a real-valued function $f$ is the subset of the domain containing the elements which are not mapped to zero. For a more formal definition the reader is addressed to \cite{Rudin_principles1953}. 
}. For the values of $m_g$ for which the value of the prior $\rho$ is not zero, the prior is then a multiplicative constant that multiplies the likelihood to obtain the posterior. Therefore, once again, we are not interested in the value of the prior, but we are interested in the bounds of the support of the prior, since it would appear in the ratio in Eq. \eqref{AccRatio_form1}.
The third ingredient that we are going to use in the algorithm is the distribution called $q$ in Eq. \eqref{AccRatio_form1}, the density function of the proposal distribution. The proposal distribution is used to propose a new point in the space of parameters and then to see with which probability it will be accepted or not. In general $q: \mathbb{R}^2 \to \mathbb{R}$ is not necessarily symmetric, i.e. in general $q(x,y) \neq q(y,x)$.



%

What we try to do, in other words, is to solve an \emph{inverse problem} in which the observables are the quantities used as observations in the INPOP framework. Therefore the forward model is the full dynamics of the Solar System as modeled in the INPOP ephemerides. 


The probability distribution we want to reach (i.e. the target distribution) is the posterior and it depends upon the experimental observations. We know that the posterior distribution is given by 
\begin{equation}\label{posterior_full}
\pi(m_g | \mathbf{d}_{\text{obs}}) = \frac{\rho(m_g) L(m_g | \mathbf{d}_{\text{obs}}) }{ \textstyle \int \rho(m) L(m | \mathbf{d}_{\text{obs}}) dm}
\end{equation}

The function $x \longmapsto f(x) $ of Eq. \eqref{AccRatio_form1} in our case is $m_g \longmapsto \pi(m_g | \mathbf{d}_{\text{obs}}) $.


The prior probability we decided to put on $m_g$ is a uniform probability on an interval, in particular if we call it $\rho(m_g)$, it is defined as
\begin{equation}\label{prior_dist} 
\rho(m_g) = \Biggl\{ \begin{array}{lcl} \frac{1}{b_u-b_l} & \quad & \text{if } m_g \in [b_l, b_u] \\ 0 & \quad & \text{otherwise}  \end{array}
\end{equation} 
where $b_l$ and $b_u$ are respectively the lower bound and the upper bound to the mass. We use $b_l = 0 \; eV c^{-2} $ as lower bound and $b_u = 3.6 \times 10^{-23} eV c^{-2}$ as upper bound. The upper bound is chosen according to the threshold provided by Bernus et al. in \cite{Bernus2020}.
Regarding the likelihood, $m_g$ then is the parameter we want to sample and and $\mathbf{d}_{\text{obs}}$ the set of observations used for the costruction of the posterior. 
Although the Eq. \eqref{posterior_full} is correct, we are not interested \emph{exactly} in the posterior, but in the posterior up to a constant. 
A simpler form of Eq. \eqref{posterior_full} is \begin{equation}\label{posterior_form2}
\pi(m_g | \mathbf{d}_{\text{obs}}) \propto \rho(m_g) L(m_g | \mathbf{d}_{\text{obs}}) 
\end{equation}
since the denominator is a constant and so we can forget it. In Eq. \eqref{posterior_form2} $L(m_g | \mathbf{d}_{\text{obs}})$ is the likelihood function.

Following Eq. (3) of \cite{Mosegaard1995MonteCS} to solve an inverse problem of the kind 
 \begin{equation}  
\mathbf{g}(m_g)= \mathbf{d}_{\text{obs}}
 \end{equation}
if we describe experimental results by a vector of observed values $\mathbf{d}_{\text{obs}}$  with independent, identically distributed Gaussian uncertainties described by a covariance matrix $\mathbf{C}=diag(\sigma_i^2)$, then
(see Eq. (15) and (16) of \cite{Mosegaard1995MonteCS})

\begin{equation}\label{likelihood_form1}
\scalebox{0.9}{ $ L(m_g | \mathbf{d}_{\text{obs}}) \propto \exp \left( -\frac{1}{2} \sum_i \left( \frac{g^i(m_g) - d^i_{\text{obs}}}{\sigma_i} \right)^2 \right).$}
\end{equation}

In Eq. \eqref{likelihood_form1} we have the misfit function  that in our specific INPOP case is $\chi^2$ as explained in  Sec. \ref{sec_Differences_standard_approach}.
For sake of simplicity we can rewrite Eq. \eqref{likelihood_form1} as 
\begin{equation}\label{likelihood_form2}
L(m_g | \mathbf{d}_{\text{obs}}) \propto \exp \left( -\frac{1}{2} \chi^2(m_g) \right)
\end{equation}
and then in our graviton case Eq. \eqref{posterior_form2} becomes
\begin{equation}\label{posterior_v3}
\pi(m_g | \mathbf{d}_{\text{obs}}) \propto \rho(m_g)  \exp \left( -\frac{1}{2} \chi^2(m_g) \right).
\end{equation}

Therefore in particular if $m_g \notin [b_l, b_u]$ (lower and upper bound for $m_g$) then $\rho(m_g)=0$ by definition of our prior. The observations are considered to be Gaussian variables
so prior and likelihood are choosen in some sense \emph{independently}. The acceptance ratio, according to the MH algorithm rule given in Eq. \eqref{AccRatio_form1}, assuming $m^1_g$ as an already accepted value and $m^2_g$ as a proposed value, is 

\begin{widetext}
\begin{equation}\label{AccRatio_form2}
\om{
p_a(m_g^1, m_g^2) = \Biggl\{ \begin{array}{lcl}  \min \left( 1, \frac{ \pi(m^2_g | \mathbf{d}_{\text{obs}}) q(m_g^2 | m_g^1 ) }{ \pi(m^1_g | \mathbf{d}_{\text{obs}}) q(m_g^1 | m_g^2 ) } \right) & \quad & \text{if } m^2_g \in [b_l, b_u] \\ 0 & \quad & \text{otherwise}  \end{array} 
}
\end{equation} 
\end{widetext}
 From now on we will substitute the symbol $p_a(m_g^1, m_g^2)$ with $\alpha(m_g^2 | m_g^1)$. 

\subsubsection{Key differences with respect to the standard MCMC approach}\label{sec_Differences_standard_approach}

In the algorithm we used for our simulations, there are two key differences with respect to the
classical MH algorithm presented in the literature:
\begin{itemize}
\item We do not provide \emph{directly} the forward problem to the algorithm  (i.e. we
do not use $\mathbf{g}$) as expressed in Eq. \eqref{likelihood_form1}  since our forward problem \emph{is} the full INPOP ephemerides construction process (see Sect. \ref{sec:inpop}) and it is too computationally expansive to use directly in the MH algorithm. The MH algorithm is intrinsically sequential and just one run of $\mathbf{g}(m_g)$ would take  up to 8 hours. 
\item  Instead we provide an estimation of the \emph{normalized} $\chi^2(m_g, \mathbf{k})$
where 
$m_g$ is the mass of the graviton $m_g$ for which we want to compute the corresponding $\chi^2$ value and $\mathbf{k}$ are all the other astronomical INPOP parameters.
\end{itemize}
In particular we have
\begin{equation}\label{normalized_chi2_form1}
\scalebox{0.95}{$ \chi^2(m_g, \mathbf{k}) \equiv \frac{1}{N_{\text{obs}}} \sum_{i=1}^{N_{\text{obs}}} \left( \frac{g^i(m_g, \mathbf{k}) - d^i_{\text{obs}}}{\sigma_i} \right)^2$ }
\end{equation}
used in Eq. \eqref{likelihood_form1} and where $N_{\text{obs}}$ is the number of observations. 
For simplicity of notation we define $\chi^2(m_g)$ such as
\begin{equation}\label{normalized_chi2_form2}
\chi^2(m_g) \equiv  \chi^2(m_g, \mathbf{k}).
\end{equation}
Given the definition of  $\chi^2(m_g)$ we consider the function $m_g \longmapsto \chi^2(m_g)$ estimated as follows:
we call  $\mathbf{k}$ the values of the astronomical parameters obtained after fit for a given fixed $m_g$ and then we obtain as output $\chi^2(m_g)=\chi^2(m_g, \mathbf{k})$.
As it is time consuming to compute the full forward problem for each value of $m_g$, we compute an approximation of $\chi^2(m_g)$ (Sect. \ref{sec:GP_and_lin_interp}).
From now on we call the approximation used as $\tilde{\chi}^2(m_g)$ and we will specify better what $\tilde{\chi}^2(m_g)$ is, later on, depending on the case. We assume for sake of simplicity that \[ \chi^2(m_g) \approx \tilde{\chi}^2(m_g) \quad \forall m_g \] whatever the approximation of the normalized $\chi-$squared $\chi^2(m_g)$ is (linear interploation or Gaussian Process).
Assuming so, the approximation $\tilde{\chi}^2(m_g)$ yields to an approximated version of the likelihood $L(m_g | \mathbf{d}_{\text{obs}} )$ up to a constant factor.
We call 
\begin{equation}\label{approx_likelihood_form1}
 \tilde{L}(m_g | \mathbf{d}_{\text{obs}} ) \equiv \exp \left( - \frac{N_{\text{obs}}}{2} \tilde{\chi}^2(m_g) \right).
\end{equation}
From now the "likelihood" is actually the form of the approximated likelihood given in Eq. \eqref{approx_likelihood_form1}, unless differently specified.

In Eq. \eqref{approx_likelihood_form1} we multiplied $\tilde{\chi}^2(m_g)$ by $N_{\text{obs}}$. This is necessary considering that 
\[ L(m_g | \mathbf{d}_{\text{obs}}) \propto  \exp \left( -\frac{N_{\text{obs}}}{2} \chi^2(m_g) \right) \] and \[ \exp \left( -\frac{N_{\text{obs}}}{2} \chi^2(m_g) \right) \approx  \exp \left( -\frac{N_{\text{obs}}}{2} \tilde{\chi}^2(m_g) \right) \] thus \[ L(m_g | \mathbf{d}_{\text{obs}}) \approx \tilde{L}(m_g | \mathbf{d}_{\text{obs}} ). \] 
As in the MH algorithm we don’t use the actual likelihood, but an approximation of the likelihood, we then write an approximation of the probability density function of the posterior $\tilde{\pi}(m_g | \mathbf{d}_{\text{obs}} )$ as (up to a constant factor)
\[
\begin{split}
\tilde{\pi}(m_g | \mathbf{d}_{\text{obs}} ) & \propto \rho(m_g)\tilde{L}(m_g | \mathbf{d}_{\text{obs}}) = \\
 & = \rho(m_g)\exp \left( -\frac{N_{\text{obs}}}{2} \tilde{\chi}^2(m_g) \right)  .
\end{split}
\]
Therefore we can consider the posterior probability distribution as
\begin{equation}\label{posterior_form3}
 \tilde{\pi}(m_g | \mathbf{d}_{\text{obs}} ) \propto \rho(m_g)\exp \left( -\frac{N_{\text{obs}}}{2} \tilde{\chi}^2(m_g) \right)
\end{equation} 
and each time we are going to refer to the posterior, we will actually speak about Eq. \eqref{posterior_form3}.
 
 $N_{\text{obs}}$ is a known integer number, therefore given $m_g$ to compute $\tilde{\pi}(m_g | \mathbf{d}_{\text{obs}})$ we just need to compute $\tilde{\chi}^2(m_g)$. Moreover, the point of using an approximation is exactly here: we have an easy (but more importantly, fast) way to compute $\tilde{\chi}^2(m_g)$.

\subsubsection{Implementation}
We assume that we are building our Markov Chain and we have accepted $\overline{m}_g$ as the last mass in the chain. We build the next value of the mass in the chain as follows.

\begin{enumerate}
\item We propose a new value of the mass $m^*_g \sim \mathcal{N}(\overline{m}_g, \sigma_1^2)$ where $\sigma_1$ is a fixed standard deviation we have decided. The probability density of proposal $q$ will be

\begin{equation}\label{eq:GaussianProposal} \small
q(m^*_g | \overline{m}_g) = \frac{1}{\sigma_1 \sqrt{2 \pi}} \exp \left( - \frac{1}{2} \left( \frac{m^*_g - \overline{m}_g}{\sigma_1} \right)^2 \right).
\end{equation}
Let's note that \[ q(m^*_g | \overline{m}_g) = q(\overline{m}_g | m^*_g). \]

\item Given $m^*_g$ we compute $\tilde{\chi}^2(m_g^*)$.

\item In the end we compute the acceptance ratio for $m^*$ given $\overline{m}_g$, that for our MH algorithm becomes:

\begin{widetext}
\om{\begin{equation}\label{AccRatio_form3} 
\alpha(m^*_g | \overline{m}_g) = \Biggl\{ \begin{array}{lcl} \min \left( 1, \frac{ \tilde{\pi} (m^*_g | \mathbf{d}_{\text{obs}}) q(m_g^* | \overline{m}_g )  }{ \tilde{\pi} ( \overline{m}_g | \mathbf{d}_{\text{obs}}) q(\overline{m}_g | m_g^* ) } \right) & \quad & \text{if } m^*_g \in [b_l, b_u] \\ 0 & \quad & \text{otherwise} \end{array}
\end{equation}}
\end{widetext}

Starting from Eq. \eqref{AccRatio_form3} we proceed with some computations in order to clarify the final form Eq. \eqref{AccRatio_form4}, assumed by the acceptance probability $\alpha(m^*_g | \overline{m}_g)$.
Indeed, if $m_g^* \in [b_l, b_u]$ the prior holds \[ \rho(m_g^*) = \rho (\overline{m}_g) .\]
Thus we have



\begin{equation}\label{AccRatio_form5}
\begin{split}
\frac{\tilde{\pi}( m^*_g | \mathbf{d}_{\text{obs}} ) q(m^*_g | \overline{m}_g ) }{\tilde{\pi}( \overline{m}_g | \mathbf{d}_{\text{obs}} ) q(\overline{m}_g | m^*_g) } & = \frac{\tilde{\pi}( m^*_g | \mathbf{d}_{\text{obs}} ) }{\tilde{\pi}( \overline{m}_g | \mathbf{d}_{\text{obs}} ) } \\
& =  \frac{\rho(m_g^*) \tilde{L} (m^*_g | \mathbf{d}_{\text{obs}}) }{\rho(\overline{m}_g) \tilde{L} (\overline{m}_g | \mathbf{d}_{\text{obs}}) } \\
& = \frac{ \tilde{L} (m^*_g | \mathbf{d}_{\text{obs}}) }{  \tilde{L} (\overline{m}_g | \mathbf{d}_{\text{obs}}) }
\end{split}
\end{equation}

 since \[ q(m^*_g | \overline{m}_g) = q(\overline{m}_g | m^*_g) \] and $\rho(m_g^*) = \rho (\overline{m}_g)$.

Eq. \eqref{AccRatio_form5} also implies that if $ m^*_g \in [b_l, b_u] $ then thanks to Eq. \eqref{approx_likelihood_form1}


{\begin{eqnarray}\label{AccRatio_form4}
\scriptsize
&&\alpha(m^*_g | \overline{m}_g) \\
&&= \min \left( 1, \exp \left( - \frac{N_{\text{obs}}}{2} \left( \tilde{\chi}^2(m^*_g) - \tilde{\chi}^2(\overline{m}_g)   \right) \right)\right). \nonumber 
\end{eqnarray}}

At the light of the Eq. \eqref{AccRatio_form5} it is clear that we are actually using the so called Metropolis algorithm and not the MH algorithm.

\end{enumerate}

\subsubsection{Gaussian Process and linear interpolation} \label{sec:GP_and_lin_interp}


Generally speaking we approximate the function $m_g \longmapsto \tilde{\chi}^2 (m_g)$ considering a set of values $ S = \{ \left( \overrightarrow{m}_g, \chi^2(\overrightarrow{m}_g) \right) \} $ and interpolating among the points of the set $S$. $\overrightarrow{m}_g$ gathers values of the masses for which we are going to compute the actual corresponding $\chi^2$ \vm{unapproximated}.




The interpolation $m_g \longmapsto \tilde{\chi}^2 (m_g)$ has been built in exploiting a Gaussian Process among the points of the set $S$.
The Gaussian Process Regression (GPR) is a method to predict a continuous variable as a function of one or more dependent variables, where the prediction takes the form of a probability distribution (see e.g. \cite{Strub:2022upl, rasmussen2005gaussian}). A Gaussian Process is a family of random variables $h(x)$, indexed by a set $x \in X$, and with assuming joint Gaussian distribution
(see \cite{rasmussen2005gaussian} for further details).
In our case the $h$ will be $m_g \longmapsto \tilde{\chi}^2(m_g)$ and the GPR will be applied to some $\chi^2$ values estimated with INPOP for a given values of $m_g$.
A Gaussian Process is specified by a given mean function $m_f: X \to \mathbb{R}$, a covariance function $K(x, x') : X \times X \to \mathbb{R}$ and a so called \emph{training set} that we denote as $X \supsetneq X_t = \{ x_1, \dots, x_n \}$. 
In particular it holds that
\[ h(X_t) = \begin{bmatrix} h(x_1) \\ \vdots \\ h(x_n) \end{bmatrix} \sim \mathcal{N}(m_f(X_t), K(X_t, X_t)), \]
where 
\[  m_f(X_t) =  \begin{bmatrix} m(x_1) \\ \vdots \\ m(x_n) \end{bmatrix} \]
and
\begin{equation}\label{GP_kernel_eq2}
 K(X_t, X_t) = 
  \begin{pmatrix}
    K(x_1, x_1) & \dots & K(x_1, x_n) \\
    \vdots & \vdots & \vdots \\
    K(x_n, x_1) & \dots & K(x_n, x_n)
  \end{pmatrix} 
\end{equation}
We assume noise-free observations $\mathbf{d}=h(X_t)$ and we can estimate the values $h(X_p)$  at the points $X_p \subseteq X$ as they are described by the Gaussian distributions by definition \vm{with Eq. \eqref{GPeq1}}
\begin{equation}\label{GPeq1}
\small
\begin{split}
h(X_p) & \sim \mathcal{N}(K(X_p, X_t) K(X_t, X_t)^{-1} \mathbf{d}, \\
	   & K(X_p, X_p) - K(X_p, X_t)K(X_t, X_t)^{-1} K(X_t, X_p)).
\end{split}
\end{equation}

The reader is referred to \cite{rasmussen2005gaussian} for a more detailed discussion on the Gaussian Processes and to \cite{GPfit2015} for the documentation of the package \textbf{GPfit} that we used to produce our GPR. The notation used for the general GPR description is similar to what is found in \cite{Strub:2022upl, rasmussen2005gaussian}.
%
In our case the $m$ function is the Best Linear Unbiased Predictor (BLUP), as described in \cite{GPfit2015} and \cite{Robinson1991}.
We used a standard exponential kernel such as

\[ K(x, x') = \exp \left( - \frac{|x - x'|^{\alpha}}{ \ell^2} \right) \]  for suitable $\ell$ and $\alpha$ parameters to tune. 
%
The points we are interpolating with the GPR (in GPR jargon the \emph{observations}, see \cite{rasmussen2005gaussian} ) are considered as noise-free, since they are computer simulations. We refer the reader to \cite{Sacks1989} for further details about this assumption. %
%
However, as we are using the GPR for interpolating $\chi^2$ values that will be seen as forward model outcome by the MCMC, it is interesting to consider the uncertainty of this interpolation (see \cite{Sacks1989} and \cite{Santner2003}) for the interpretation of the MCMC results.

\subsubsection{GPR and interpolation uncertainty} \label{GP_and_int_uncertainty}
One of the advantages about GPR, is that, in principle, you can consider the value of $\tilde{\chi}^2(m_g)$ as a normal random variable with given average and standard deviation provided as outcome of the GPR. Let's indicate as 
 \begin{equation}\label{grayline_form1} 
 m_g \longmapsto \breve{\chi}^2(m_g), 
 \end{equation}
one \emph{realization} of the GPR with its own uncertainty.

%
 As we said previously the estimator in our case is BLUP and we call it $m_g \longmapsto \tilde{\chi}^2(m_g)$. We would like also to exploit these confidence intervals and the fact that we consider
 
\begin{equation}\label{grayline_form2}
\forall m_g, \quad \chi^2(m_g) \sim \mathcal{N}(\tilde{\chi}^2(m_g), \tilde{\sigma}^2(m_g)).
\end{equation}

In Eq. \eqref{grayline_form2} $\tilde{\sigma}(m_g)$ is the value provided by the GPR to use as standard deviation when we consider $\chi^2(m_g)$ as a normal random variable.
We summarize in Table \ref{tab:Table_notation} what are the differences between $\chi^2(m_g), \tilde{\chi}^2(m_g)$ and $\breve{\chi}^2(m_g)$. 
We have a set $ S = \{ \left( \overrightarrow{m}_g, \chi^2(\overrightarrow{m}_g) \right) \}$ that can be seen as a set of couples of the form $\left( m, \chi^2(m) \right) \in \mathbb{R}^2$ as  
\begin{equation}\label{S_form2}
S = \{ \left( m, \chi^2(m) \right) \}_{m \in \overrightarrow{m}_g }.
\end{equation}
Starting from Eq. \eqref{S_form2}  we define the $b_l$ and $b_u$ to use in the prior as:
\begin{equation}\label{lower_bound_def}
b_l \equiv \min \biggl\{ m \; | \; \left( m, \chi^2(m) \right) \in S \biggr\},
\end{equation}

\begin{equation}\label{upper_bound_def}
b_u \equiv \max  \biggl\{ m \; | \; \left( m, \chi^2(m) \right) \in S \biggr\}.
\end{equation}
Definitions given in Eq. \eqref{lower_bound_def} and Eq. \eqref{upper_bound_def} are necessary since we cannot interpolate outside of the minimum and maximum values available for masses. 
Once that $b_l$ and $b_u$ are available, we produce a mesh of $M$ equispaced points inside the interval $ [ b_l, b_u ]$.

\begin{equation}\label{mesh_definition}
b_l= m^0_g < m^1_g < \cdots < m^M_g < m^{M+1}_g = b_u 
\end{equation}
We define then for the mass $m^i_g$ and $\forall i \in \{0, \dots, M+1 \}$, a corresponding perturbation of $\tilde{\chi}^2(m^i_g)$ that is drawn following a normal random variable with mean and standard deviation provided. Similarly $\forall i \in \{0, \dots, M+1 \}$ we call $\breve{\chi}^2_i$ such a perturbation of $\tilde{\chi}^2(m^i_g)$ that is drawn as follows
\[ \breve{\chi}^2_i \sim \mathcal{N}(\tilde{\chi}^2(m_i), \tilde{\sigma}^2(m_i)). \]

In this way we have a set 
\begin{equation}\label{single_pert_form1}
\breve{S}^{M} = \biggl\{ \left( m^i_g, \breve{\chi}^2_i \right) \biggr\}_{i=0}^{M+1}
\end{equation}
that represent a discrete version of a possible perturbation of $m_g \longmapsto \tilde{\chi}^2(m_g)$. The function
\begin{equation}\label{single_pert_form2}
    m_g \longmapsto \breve{\chi}^2(m_g)
\end{equation}
is then defined as the linear interpolation of the points given in Eq. \eqref{single_pert_form1}.
We call the form of perturbed $\chi^2$ given in Eq. \eqref{single_pert_form2} Gaussian Process Uncertainty Estimation (GPUE).
The idea  is: $m_g \longmapsto \breve{\chi}^2(m_g)$ represents a possible perturbation to the nominal interpolation that we have obtained with the GPR, in the sense that by definition $m_g \longmapsto \breve{\chi}^2(m_g)$ is locally defined as a realization of a normal random variable . We repeat the computation of Eq. \eqref{single_pert_form1} for several times.
Whereupon for each of the maps $m_g \longmapsto \breve{\chi}^2(m_g)$ produced, we run separately a Markov Chain. Doing so we are running MH algorithm to produce a Markov Chain on a noisy version of the nominal interpolation obtained with GPR. This method is an attempt to assess the \emph{uncertainty of interpolation} that we have. 
In the rest of this work, we call Gaussian Process Uncertainty Realization (GPUR) such a final posterior.

\subsubsection{Comparison between GPR and Linear Interpolation} \label{sec:GPR_LinInt_Comparison}
The first choice to interpolate points is usually the linear interpolation. The main drawback of using such a tool is the lack of information on the accuracy for the obtained approximation. In our case, we are going to show that the linear interpolation can be encompassed inside the interpolation uncertainty provided by the GPR at $2\sigma$ level. 
The upper bound for the mass $m_g$ at $99.7 \%$ confidence interval provided by Bernus et al. in \cite{Bernus2020} is $m_g \leq 3.62 \times 10^{-23} \; eV c^{-2}$, then the interval $0 \leq m_g \leq 3.62 \times 10^{-23} \; eV c^{-2}$ is the one in which to search for a more detailed analysis.

In the remaining part of the current section (Sec. \ref{sec:GPR_LinInt_Comparison}), given a set $S$ as in Eq.  \eqref{S_form2}, we indicate, respectively, with $\tilde{\chi}^2_{LI}$ and $\tilde{\chi}^2_{GP}$ the approximations obtained with a linear interpolation on the set $S$ and a GPR on the set $S$.

\begin{figure}[ht]
\centering
  \includegraphics[width=.99\linewidth]{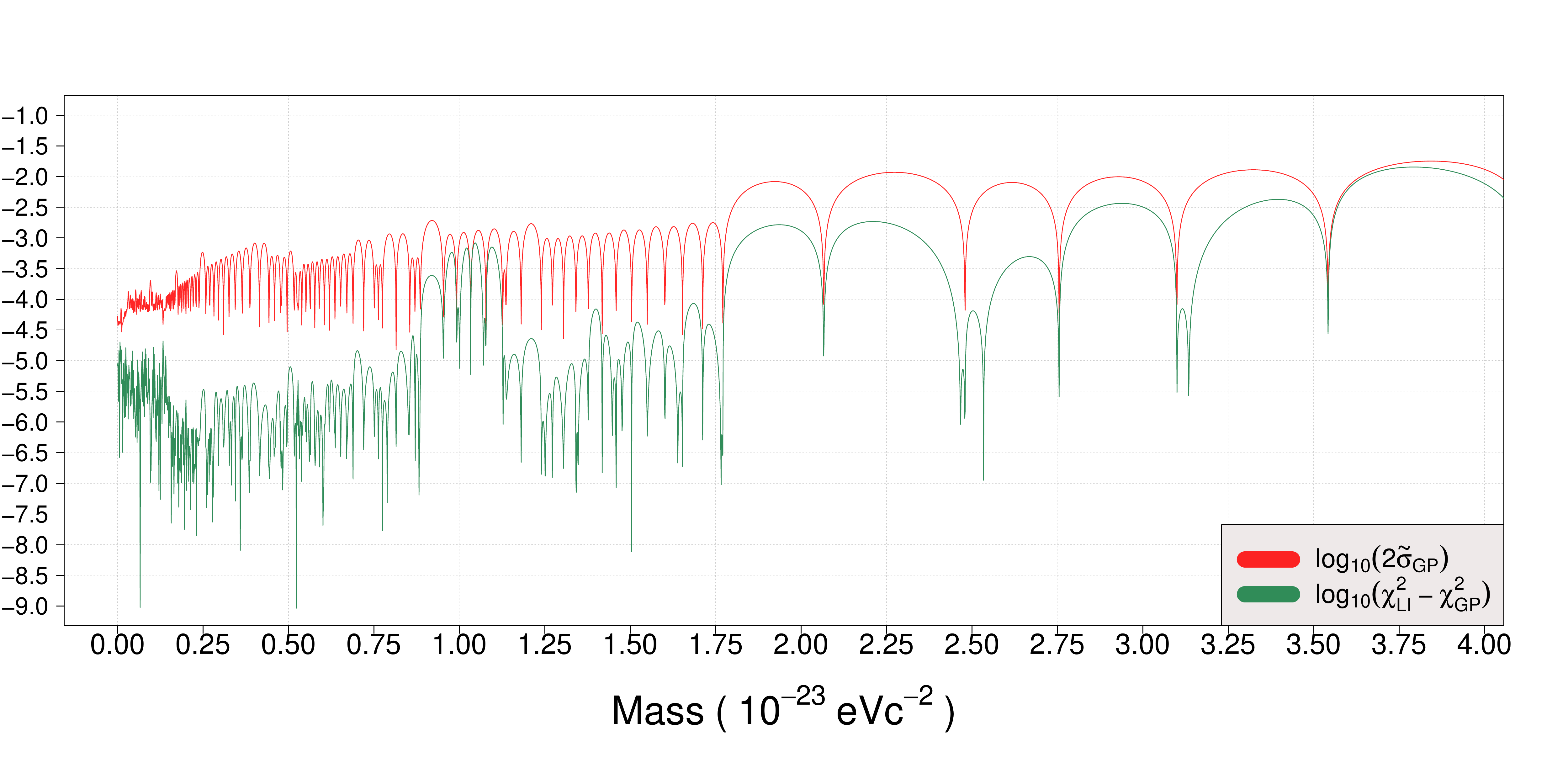}
   \caption{Log plot to compare $|\tilde{\chi}_{LI}^2(m_g) - \tilde{\chi}_{GP}^2(m_g)|$ and $2\tilde{\sigma}(m_g)$ in the interval $0 \leq m_g \leq 4.00 \times 10^{-23} \; eV c^{-2}$.}
    \label{fig:GPR_vs_LI_plot_IT55_log10}
\end{figure}

\begin{figure}[ht]
\centering
  \includegraphics[width=.99\linewidth]{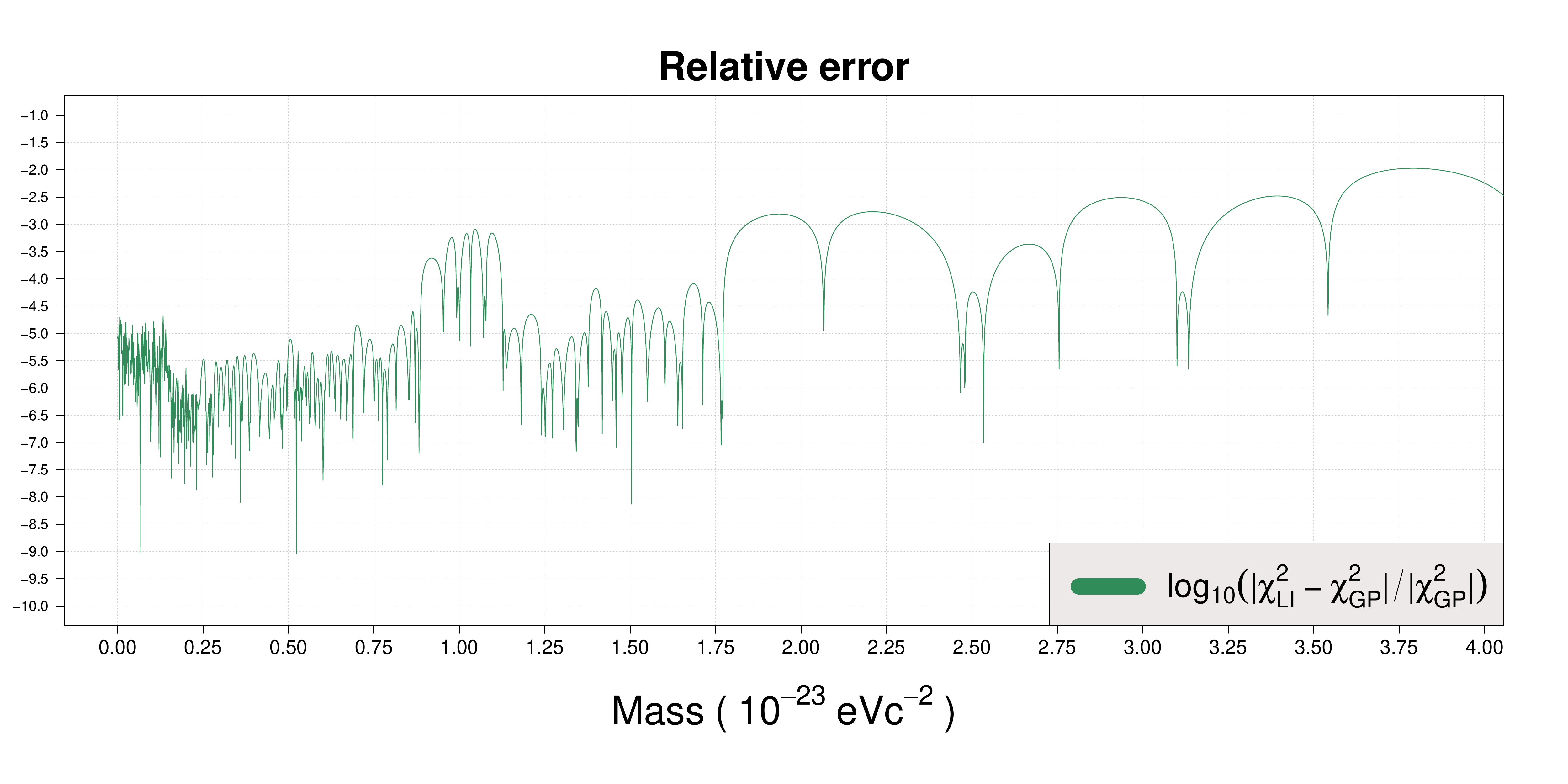}
   \caption{Relative error between $\tilde{\chi}_{LI}$ (linear interpolation) and $\tilde{\chi}_{GP}$ (GPR) for masses between $0 \leq m_g \leq 4.00 \times 10^{-23} \; eV c^{-2}$.}
    \label{fig:RelError_GPLI_full}
\end{figure}

As shown on Fig. \ref{fig:GPR_vs_LI_plot_IT55_log10}, the difference $|\tilde{\chi}_{LI}^2(m_g) - \tilde{\chi}_{GP}^2(m_g)| $ is smaller or equal to  $2 \tilde{\sigma}(m_g)$ for all values $0 \leq m_g \leq 4 \times 10^{-23} \; eV c^{-2}$. 
Therefore the linear interpolation $\tilde{\chi}_{LI}$ is encompassed inside the $2\sigma$ uncertainty on $\tilde{\chi}_{GP}$ provided by the GPR. 
On Fig. \ref{fig:RelError_GPLI_full} is shown the relative error of $\tilde{\chi}_{LI}^2(m_g)$ with respect to $\tilde{\chi}_{GP}^2(m_g)$ for the set $S$. For $m_g$ close to $3.75 \times 10^{-23} \; eV c^{-2}$ the relative error is close to $0.01$, whereas for $m_g \leq 3.50 \times 10^{-23} \; eV c^{-2}$ the relative error is smaller than $0.0032$. Considering instead values $0 \leq m_g \leq 1.75 \times 10^{-23} \; eV c^{-2}$ the relative error is below $0.001$ (see Fig. \ref{fig:RelError_GPLI_full}). 

Finally, since  $|\tilde{\chi}_{LI}^2(m_g) - \tilde{\chi}_{GP}^2(m_g)| \leq 2\tilde{\sigma}(m_g)$, in light of the methodology proposed in Sec. \ref{GP_and_int_uncertainty}, it is not necessary to run MCMC for $\tilde{\chi}^2 = \tilde{\chi}^2_{LI} $. Indeed proceeding as described in Sec. \ref{GP_and_int_uncertainty}, $\tilde{\chi}^2_{GP}(m_g)$ is perturbed as a normal random variable of standard deviation $\tilde{\sigma}(m_g)$ in which is then included also the case $\tilde{\chi}_{LI}^2(m_g)$ . The linear case $ \tilde{\chi}^2 = \tilde{\chi}_{LI}^2 $ is then encompassed within the analysis of the GPR uncertainty.

\subsubsection{Bayes Factor} \label{sec:BF_theory}

The Bayes factor is a tool used in the context of model selection (see, e.g. \cite{GillJeff2008}). The central notion is that prior and posterior information should be combined in a ratio that provides evidence of one model specification $M_1$ over an other $M_2$. \vm{The Bayes Factor can be interpreted as a quantity saying which model between $M_1$ and $M_2$ represent at the best the observed data set $\mathbf{d}_{\text{obs}}$.  }
In our case selecting a specific model means to select a specific value of $m_g$.
The Bayesian setup \vm{requires} a prior unconditional distribution for the parameter $m_g$ we are dealing with, that, for us, is $\rho(m_g)$.  
The quantity of interest is then the ratio:
\begin{equation} \label{eq:BF_Definition}
BF= \frac{\pi(M_1 | \mathbf{d}_{\text{obs}} )}{\pi(M_2 | \mathbf{d}_{\text{obs}} )} \times \frac{\rho(M_2)}{\rho(M_1)}.
\end{equation}
In the case of equal priors for $M_1$ and $M_2$ the Bayes Factor equals to the ratio of the likelihoods. For major details about the Bayes Factor and a more extensive explanation the reader is addressed to \cite{GillJeff2008}.
The outcome obtained from the Bayes Factor will be given later on Sec. \ref{sec:BF_results_it5} and \ref{sec:BF_results_it5}.
The interpretation of the Bayes Factor is not always easy. We will take as reference the so called Jeffreys' scale.
The Jeffreys' scale is used to evaluate the Bayes Factor, that is a positive number (see Eq. \eqref{eq:BF_Definition}), a qualitative judgment on the evidence. Generally speaking, if $BF \gg 1$ it indicates that the weight of Model 1 is greater than the Model 2, and on the other hand if $BF \ll 1$ then Model 2 has more weight. In the zones of the domain for which $BF \simeq 1$ neither Model 1 or Model 2 is predominant. 

\section{Results} \label{sec:Results}

\subsection{Practical implementation of MCMC diagnosis}\label{sec:MCMC_diagnosis}

%
 The strategy we used to monitor MCMC convergence follows what is suggested in Chapter 6 of \cite{brooks2011handbook}. \cite{brooks2011handbook} recommends to run different chains in parallel with different starting points. Whereupon the first part of the simulations has to be discarded (this part is called \emph{burn-in}) and so within-chains analysis is performed to monitor convergence and mixing. Once that approximate convergence is reached, all the simulations from the second halves of the chains are mixed together to summarize the target distribution. 
Proceeding in this way, there is no longer need to account for autocorrelations in the chains (see \cite{brooks2011handbook} for further details). Moreover we use the Gelman-Rubin ratio $\hat{R}$ to see if the chains have reached convergence or not.
In case $\hat{R}$ is not close to $1$, it is necessary to increase the number of steps of the chains. Besides the number of steps, it could be also necessary to tune better $\sigma_1$ as presented in Eq. \eqref{eq:GaussianProposal}.
\cite{gelmanbda04} recommends terminating simulation when at most $\hat{R} < 1.1$, although the smaller $\hat{R}-1$ is, the better the chains will have reached convergence. 
In particular we ran five chains in parallel for each MCMC process. As burn-in period we discard the first half of each of the five chains produced (see \cite{brooks2011handbook}). 
The starting value of each chain is chosen in the domain of the prior to speed up convergence towards the zone of domain of major probability. The posterior is then obtained mixing up all the five chains.

\subsection{Results obtained with INPOP ephemerides after adjustment} \label{sec:Results_it55}

The computation of $\chi^2(m_g)$ is the output, for a given value of $m_g$, of the INPOP fit after adjustment. 
We set the value of $m_g$ {to be fixed,} and proceed with the least {square adjustments} for all the other parameters.
We show the results obtained with the INPOP ephemerides after adjustment. This solution benefits {from} the full observational accuracy provided {by} modern ephemerides. First, we see the values of the Bayes Factor in Sec. \ref{sec:BF_results_it55}. Second, the outcome of the MCMC on the GPR is presented in Sec. \ref{sec:IT_55_GPR_MCMC}. Finally, the GPUR is shown in Sec. \ref{sec:IT55_results_perturbations}.
{A validation of the method is provided in Appendix \ref{sec:Results_it5} where the Bayes Factor analysis (Sect. \ref{sec:BF_results_it5}) and the results of the MCMC with GPR algorithm (Sect. \ref{sec:IT5_Results_MCMC}) have been tested with a non-zero $m_g$ simulation. {For this simulation, we have checked that the} two approaches give results consistent with a $m_g$ positive detection (BF $<$ 0.33) and {also provide a} mass determination (posterior centered on the expected value).}

\subsubsection{Bayes Factor} \label{sec:BF_results_it55}

The goal of our simulations is to see whether a massive graviton is a theory {favored over} GRT or not. To do so we first present a previous analysis totally independent from the MCMC but based only on the INPOP runs.
The tool used is the Bayes Factor as described in Sec. \ref{sec:BF_theory} (see also, e.g., \cite{GillJeff2008} and \cite{lee_wagenmakers_2014}). 
\vm{The Bayes Factor is computed with real unapproximated $\chi^2$ and not with interpolated values.}
As Model 1, we use an extremely small value for the mass of graviton i.e. $m_g \simeq 10^{-40} \times 10^{-23} \; eVc^{-2}$; whereas as Model 2, we employ a massive graviton with $m_g$ assuming the value of one component of the vector $\overrightarrow{m_g}${, simply defined as all the values $m_g$ considered in the analysis}. 
As Model 1 we use GRT, whereas as Model 2 we employ a massive graviton with \vm{$m_g \neq 0$} assuming the value of one component of the vector $\overrightarrow{m_g}$.
\begin{figure}[ht]
    \centering
    \includegraphics[width=.99\linewidth]{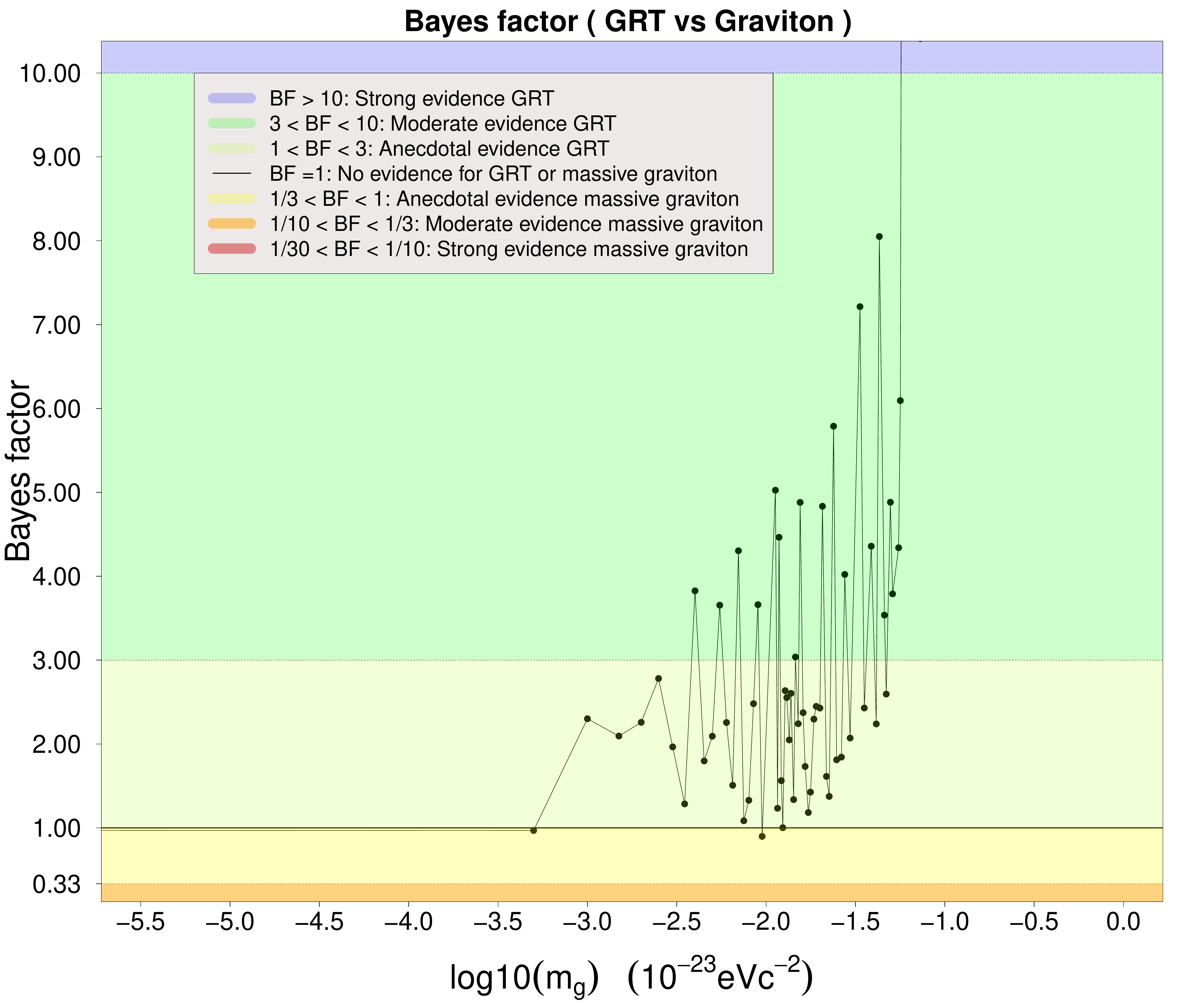}
    \caption{Value of the Bayes Factor as function of $m_g$. Each point corresponds to one INPOP run. On the $x$-axis we put the masses in terms of $\log_{10}(m_g)$. }
    \label{fig:BayesFactor55}
\end{figure}

On Fig. \ref{fig:BayesFactor55} the Bayes Factor is presented. The BF values tend to be above 1 for almost all the values of $m_g$, and it is close to 1 for values of $m_g \simeq 0$. 
The conclusion is also that GRT remains the most likely model ($BF > 10$) up to $m_g = 0.06 \times 10^{-23} \; eV c^{-2} $. For $m_g \leq 0.06 \times 10^{-23} \; eV c^{-2} $, ($BF \simeq 1$), GRT is at the same level of evidence as graviton. 
On Fig. \ref{fig:BayesFactor55} the plot is presented in the scale of $\log_{10}$ for the masses. Doing so, it is easier to see how it stays above one, and then the it tends to 1 when the $m_g$ values approach GRT.

\subsubsection{GPR and MCMC} \label{sec:IT_55_GPR_MCMC}

\begin{figure}[!ht]
\centering
  \includegraphics[width=.89\linewidth]{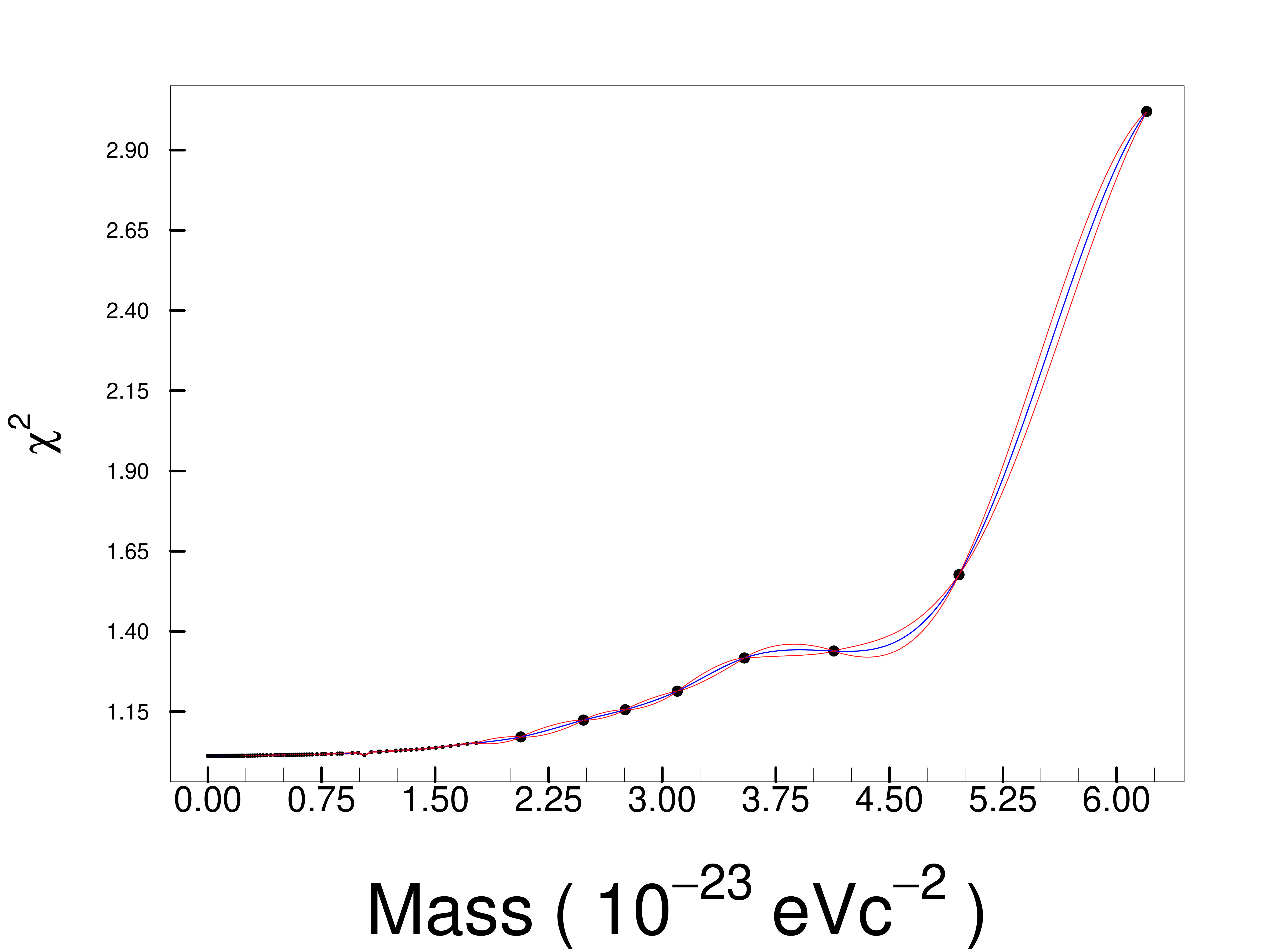}
   \caption{Plot of the function $ m_g \longmapsto \tilde{\chi}^2(m_g)$. The dots correspond to $\chi^2$ value estimated with INPOP and not extrapolated. The red lines correspond to the $2-\sigma$ uncertainties provided by the GPR.}
    \label{fig:GPR_mass_vs_chi2}
\end{figure}

\begin{figure}[!ht]
\centering
  \includegraphics[width=.99\linewidth]{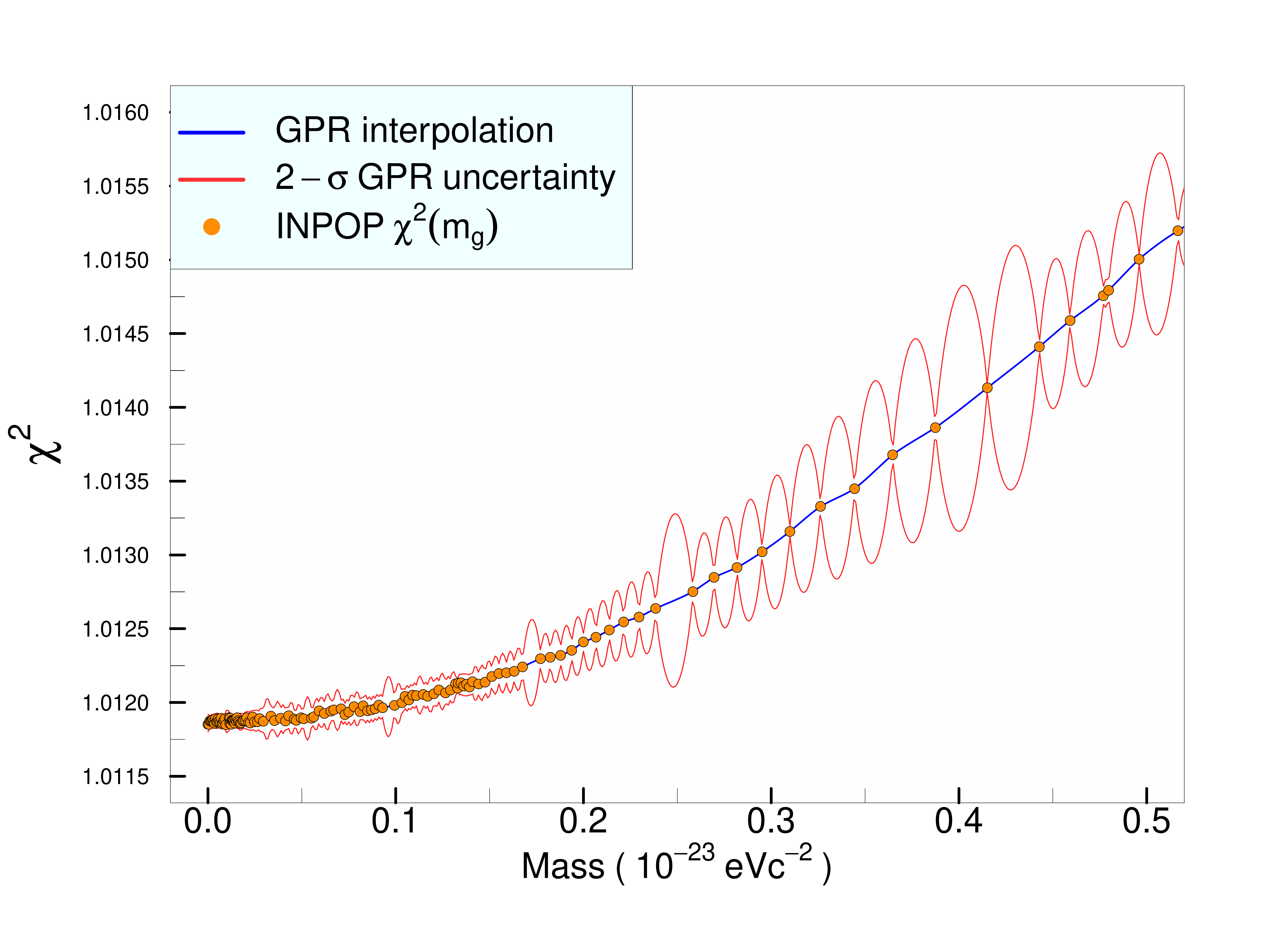}
    \caption{ Plot of the function $ m_g \longmapsto \tilde{\chi}^2(m_g)$ made up with INPOP $\chi^2$. In blue we plot the GPR interpolation among the points $\chi^2(m_g)$ computed with the full INPOP (orange dots). The red lines represent the $2-\sigma$ uncertainty (on the interpolation) provided by the GPR. }
    \label{fig:GPR_mass_vs_chi2_zoom3}
\end{figure}

On Fig. \ref{fig:GPR_mass_vs_chi2} we see the outcome of GPR on a set $S$ computed after INPOP adjustment. The blue line represents the mean value obtained from GPR (see details in Sec. \ref{sec:GP_and_lin_interp}). The red lines represent the estimates of uncertainty provided by the GPR at $2 \sigma$ level. In particular the graphs of the two red lines are $m_g \longmapsto \tilde{\chi}^2(m_g) \pm 2\tilde{\sigma}(m_g) $. We see that $m_g \longmapsto \tilde{\sigma}(m_g)$ is zero or close to zero when $m_g$ is used to compute $\chi^2(m_g)$ from INPOP. This is due to the assumption that the $\chi^2( \overrightarrow{m}_g )$ are computed with zero noise (as already pointed out in Sec. \ref{sec:GP_and_lin_interp}) as these values are direclty obtained from INPOP construction. 
With a closer look (see Fig. \ref{fig:GPR_mass_vs_chi2_zoom3} as a zoom of Fig. \ref{fig:GPR_mass_vs_chi2}  ) it appears that the uncertainty is not exactly zero at the reference points because of numerical noise necessary during the inversion (e.g. in Eq. \eqref{GPeq1}) of the covariance matrix $K$ (see Sec. \ref{sec:GP_and_lin_interp} ).

\begin{figure}[ht]
\centering
  \includegraphics[width=.99\linewidth]{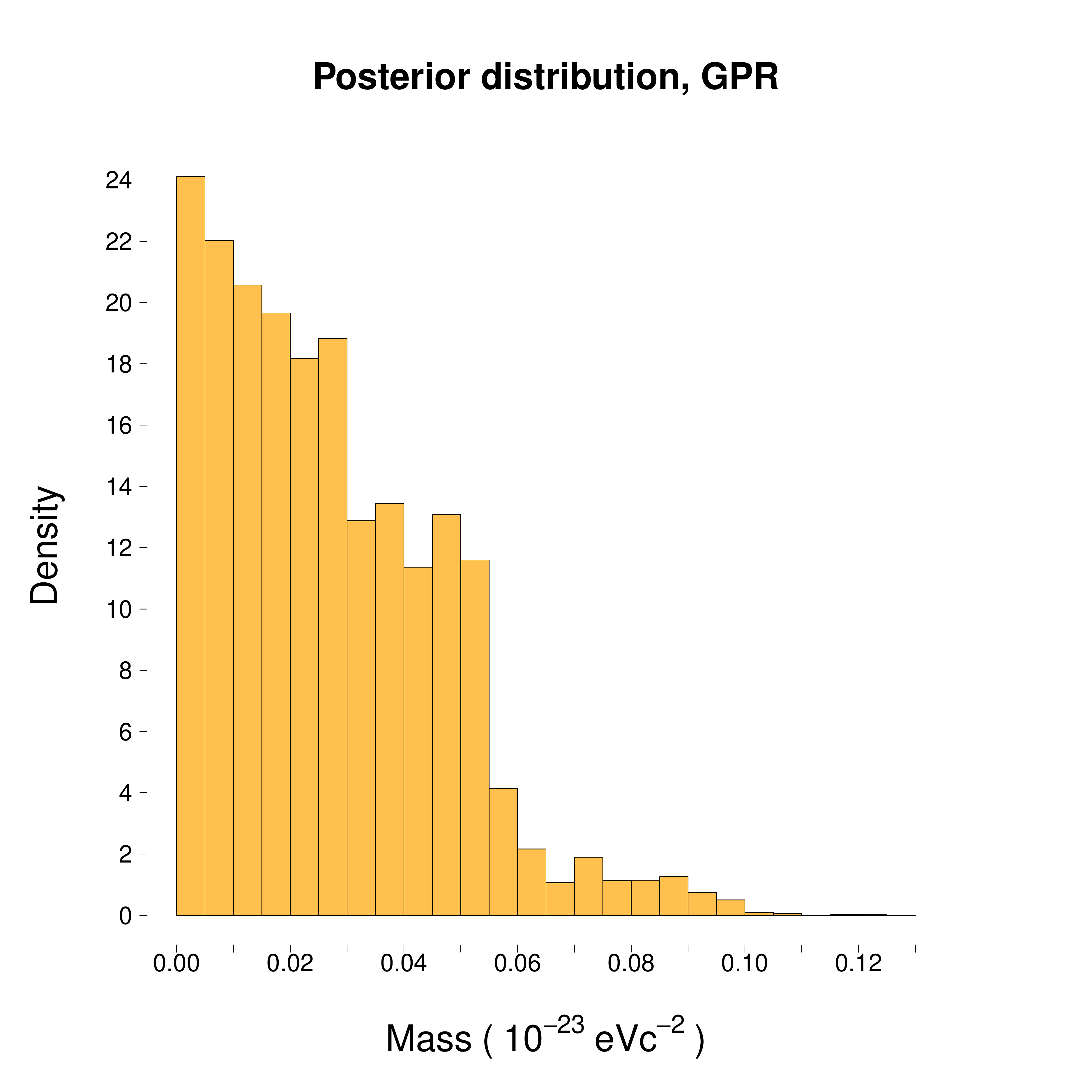}
   \caption{Density for the posterior probability distribution used as target probability.}
    \label{fig:55_iter_GPR_histogram_joint}
\end{figure}

On Fig. \ref{fig:55_iter_GPR_histogram_joint} the posterior density of probability obtained at MCMC convergence is proposed. The Gelman-Rubin factor is $\hat{R}=1.000928$. The acceptance rate is $36.9 \%$ in average among the five chains. 
For each chain we used $3 \times 10^4$ steps of the MH algorithm and a burn-in period of half chain as proposed in \cite{brooks2011handbook}. The final posterior of each chain is then built up with all the occurrences of the second half of the chains. The posterior presented on Fig. \ref{fig:55_iter_GPR_histogram_joint} is built up joining the such posteriors.
In particular Fig.  \ref{fig:55_iter_GPR_histogram_joint} can be seen as an average of the 5 densities. 
From this plot we do not have a single figure that we could choose as value for the mass of the graviton. Indeed the shape is not a \emph{bell}{, nor does it show} an individual peak. 
The quantile at $97 \%$ is $0.0985 \times 10^{-23} \; eVc^{-2}$. 
A summary of the outcome divided per chain is shown in Table \ref{tab:GPR_55it}. In the first column we put the mass with which we start the specific chain labeled $m_g^I$. In the second column we write the mean of the posterior of the single chain. This mean is labeled as $\langle m_g \rangle$. In the third column there is the acceptance rate for the single chain, after burn-in. In the fifth column the quantile of the posterior density at $99.7 \%$ is computed.
The convergence of the five chains is discussed in the Appendix \ref{app:sec:results_it_55}.
From the summary in Table \ref{tab:GPR_55it} we see that the results are consistent among the chains and tend all towards a model without massive graviton or with a mass in any case bounded by $m_g < 0.14 \times 10^{-23} \; eVc^{-2}$.
Summing up all the analyses and results obtained with our simulations (posterior density, acceptance rate, $\hat{R}$, traces of the chains, ACF plots, $m_g^I$, $0.997$-quantile, $\langle m_g \rangle$) we can say that the chains converged and that we ran enough steps for each chain.
Let us note also that the Bayes Factor presented on Fig. \ref{fig:BayesFactor55} is consistent with the MCMC outcome. 

%
%

\begin{center}
\begin{table}[!ht]
\centering
\caption{\label{tab:GPR_55it} Summary of the outcome divided by chain. $\tilde{\chi}^2$ taken as GPR; $S$ built with $\chi^2$ after INPOP fit adjustment. The unit for $m^I_g $, $ \langle m_g \rangle $ and for the $99.7 \%$ quantile is $10^{-23} eVc^{-2}$.}
\begin{tabular}{| c || c | c | c | c | c | c | }
\hline
 Chain &  $m^I_g $ & $ \langle m_g \rangle $ & Acc. rate & $99.7 \%$ quantile  \\ [1.2ex]
\hline \hline 

1 & $0.025 $ &  $ 0.0261 $ & $36.69 \%$ &  $0.094 $  \\ \hline
2 & $0.070 $ &  $ 0.0263 $ & $37.51 \%$ &  $0.090 $  \\ \hline
3 & $0.043 $ &  $ 0.0274 $ & $37.66 \%$ &  $0.094 $  \\ \hline
4 & $0.045 $ &  $ 0.0270 $ & $36.85 \%$ &  $0.097 $  \\ \hline
5 & $0.037 $ &  $ 0.0278 $ & $37.23 \%$ &  $0.098 $  \\ \hline

\end{tabular}
\end{table}
\end{center}

\subsubsection{Uncertainty quantification of interpolation errors} \label{sec:IT55_results_perturbations}
\begin{figure}[!ht]
\centering
  \includegraphics[width=.99\linewidth]{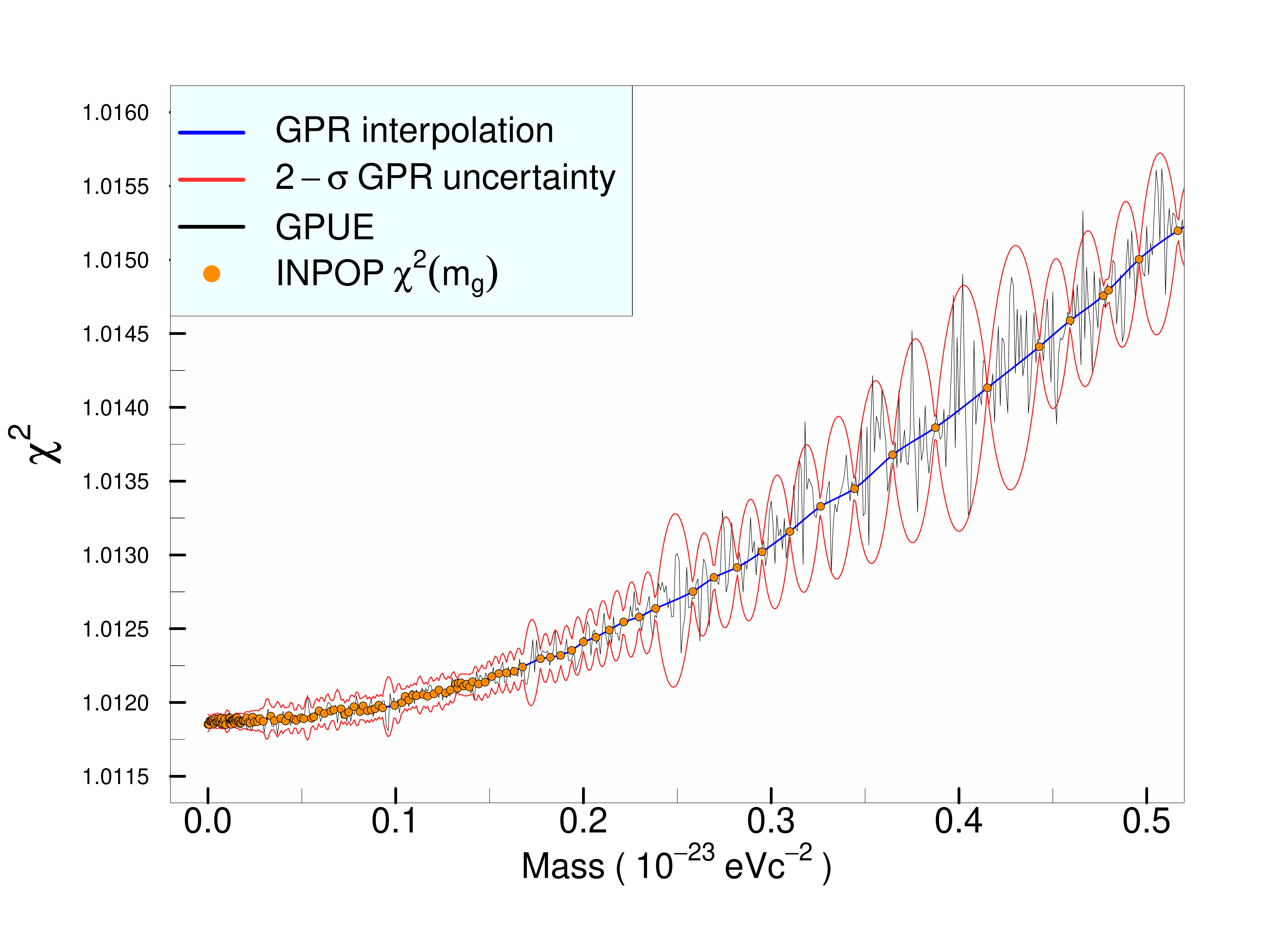}
    \caption{ Plot of the function $ m_g \longmapsto \tilde{\chi}^2(m_g)$ made up with INPOP $\chi^2$ (orange dots). In blue we plot the GPR interpolation among the points $\chi^2(m_g)$ computed with the full INPOP (orange dots). The red lines represent the $2-\sigma$ uncertainty (on the interpolation) provided by the GPR. The black like is one possible GPUE (in other words, the GPR interpolation perturbed at each point of the domain). The plot is zoomed on a portion of the domain.}
    \label{fig:GPR_only_plot_IT55_zoom3_nogrid_pert}
\end{figure}

\begin{figure}[!ht]
    \centering
    \includegraphics[width=.99\linewidth]{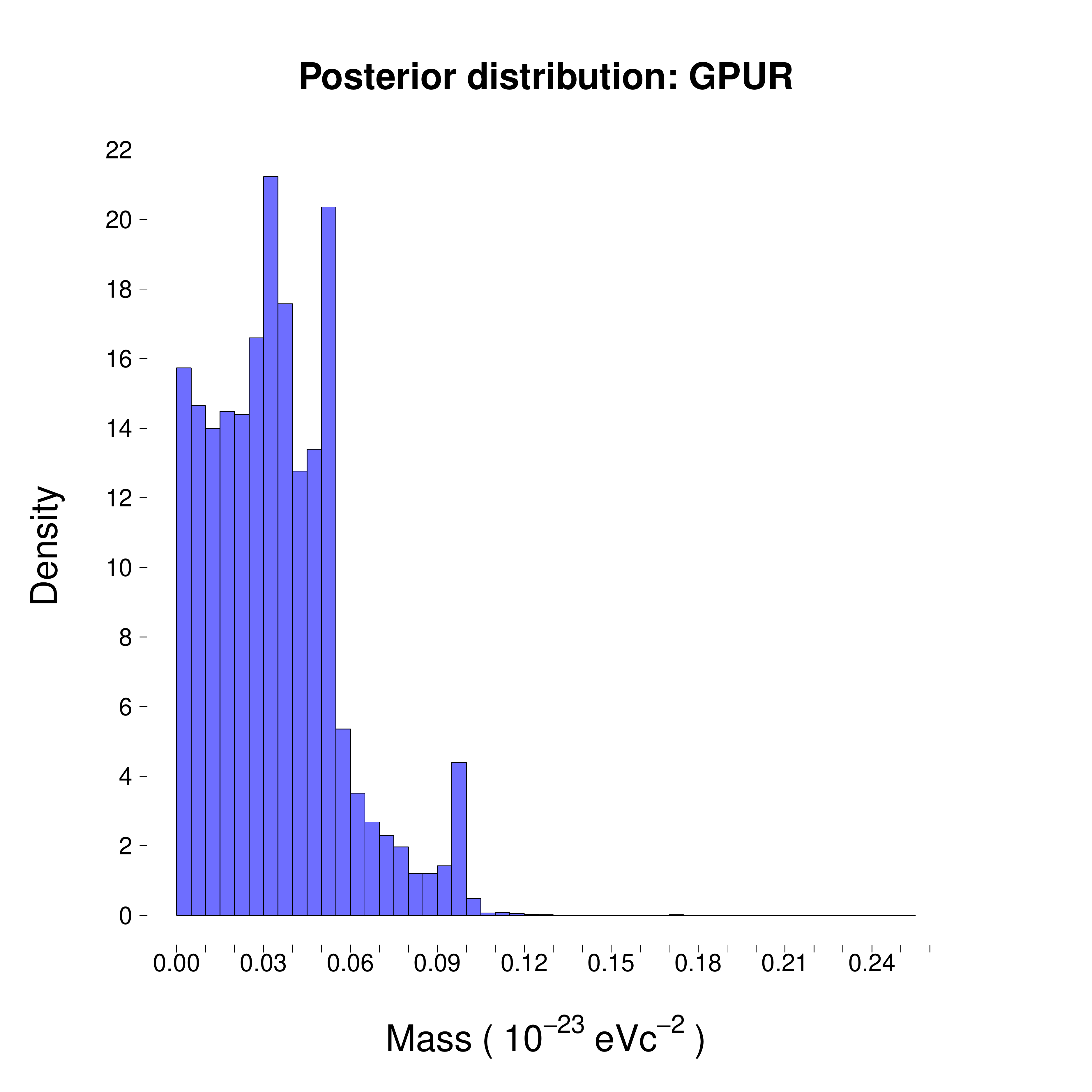}
    \caption{Density for the posterior probability distributions obtained as GPUR (blue).}
    \label{fig:55_iter_300Pert_histogram_joint}
\end{figure}

%
As explained in Sec. \ref{sec:GPR_LinInt_Comparison}, the results obtained with linear interpolation are encompassed within range of the GPR uncertainties. 
In order to explore the space of the uncertainties induced by the GPR we ran 300 MCMC processes. Each MCMC is performed on a different realization $\breve{\chi}^2$ of the GPR (Sec. \ref{GP_and_int_uncertainty}) as built in Eq. \eqref{single_pert_form2}, i.e. on a different GPUE (see Sec. \ref{GP_and_int_uncertainty}). On Fig. \ref{fig:GPR_only_plot_IT55_zoom3_nogrid_pert} we show an example of one GPUE for sake of clarity. Each MCMC process is composed from five chains in order to compute $\hat{R}$ and to have an indication of convergence. We then filtered them such that the $\hat{R}$ relative to each MCMC was properly close to $1$. The major difference we meet in the posterior with the uncertainty is the value of the mean of our histogram. It is indeed slightly greater that in the nominal case without uncertainty. On Fig.  \ref{fig:55_iter_300Pert_histogram_joint} we see the posterior obtained in including the perturbations. To summarize the results we plot together the output of all the 300 MCMC we ran. We see that the shape presented is slightly different from the one we obtained in the nominal case (GPR). It has a flat behaviour {for small values of $m_g$} except for a couple of single peaks. Actually this output is something we expected: the idea of perturbing the GPR is to give the chance for values of $\chi^2$ that are obtained for values of $m_g>0$ to be the point of minimum of $\breve{\chi}^2$. In this sense we see a general translation of the histogram towards greater values of $m_g$. On Fig. \ref{fig:55_iter_300Pert_histogram_joint} we plot the posteriors obtained from GPR (orange) and from its perturbations (blue) to have a better comparison of the two densities. On this figure the means, the $99.7\%-$quantile and the maximum value of the chains are also plotted. There is in all these three quantities a translation towards higher mass values, that can be understood as towards greater uncertainty. In other words, with respect to the GPR case, the introduction of the uncertainties induces a posterior density with less amplitude in height but larger interval of possible masses. It is interesting to see that the maximum value of the GPR chains is quite smaller than the maximum value of the perturbed realizations (GPUR), respectively $ 1.45 \times 10^{-24} \; eV c^{-2} $ and $2.53 \times 10^{-24} \; eV c^{-2}$. So the upper bound of the posterior is actually raised. We summarize in Table \ref{tab:GPR_300iterations_comparison_55it} these values.
Although different, we see that the outcomes from GPR and from the GPUR are quite similar. This is due to the very low uncertainty that we obtained with the GPR. Indeed although we accounted for these uncertainties together with the GPR, we did it according to the uncertainty provided. With small uncertainty, also the variations will be small. 

\begin{center}
\begin{table}[!ht]
\centering
\caption{\label{tab:GPR_300iterations_comparison_55it} Summary of the outcome for MCMC based on GPR and the posterior GPUR. These values are plotted as vertical lines on Fig. \ref{fig:55_iter_300Pert_GPR_histogram_comparison}. The unit is $10^{-24} eVc^{-2}$. The prior was a flat prior between $0$ and $3.62 \times 10^{-23} \; eVc^{-2}$.}
\begin{tabular}{| c || c | c | c | c | c | c | }
\hline
  &  $ \langle m_g \rangle $ & $99.7 \% $ quantile & $\max\{ m_g \} $  \\ [1.2ex]
\hline \hline 
 
GPR & $ 0.26 $ & $ 0.98 $ &  $1.45$  \\ \hline
GPUR & $ 0.34 $ & $ 1.01 $ &  $2.53 $  \\ \hline

\end{tabular}
\end{table}
\end{center}

\begin{figure}[!ht]
    \centering
    \includegraphics[width=.99\linewidth]{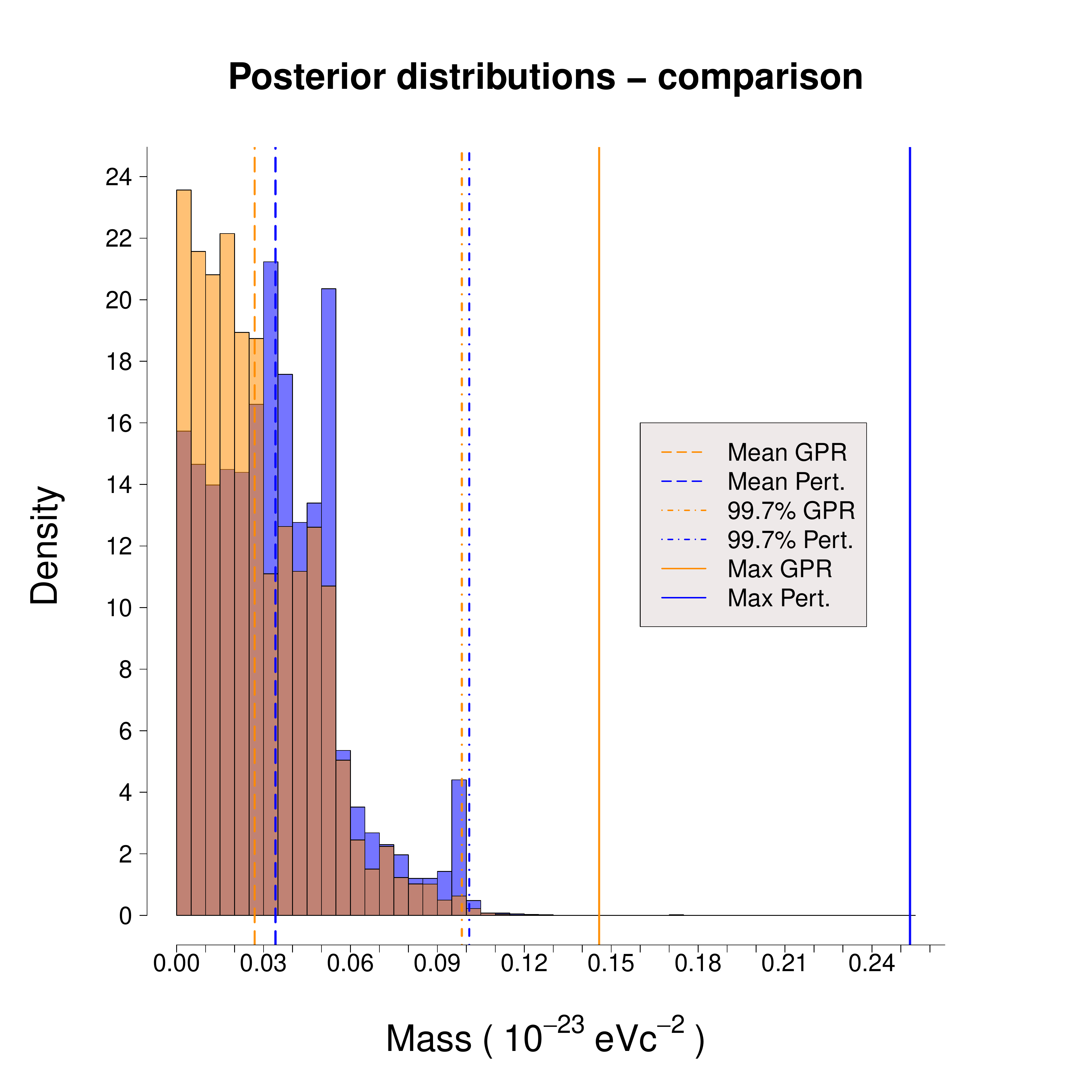}
    \caption{Densities for the posterior probability distributions obtained from GPR (in orange) and from its perturbations (blue). The dashed lines on the left represent the averages of the posterior with GPR and GPUR (respectively orange and blue). The dot-and-dashed lines represent the $99.7 \%$ quantiles of the two densities. The solid lines are instead in place of the maximum $m_g$ for each one of the two densities. The brownish area of the histogram represents the overlaid zone among the two posteriors presented.}
    \label{fig:55_iter_300Pert_GPR_histogram_comparison}
\end{figure}

\section{Discussion} \label{sec:Discussion}

\subsection{ Bayes Factor }

We have presented our work on the construction of a semi-Bayesian approach for constraining the mass of the graviton by mean of INPOP21a. Firstly, we provide a qualitative analysis using the Bayes Factor \vm{obtained with \om{an} unapproximated $\chi^2$}. It gives us the flavor of the MCMC results with respect to the final outcome of the chains. On Fig. \ref{fig:BayesFactor55} we see more evidence for GRT than for any possible ephemerides with $m_g \neq 0$, having in any case no evidence for non-zero graviton masses. The MCMC becomes then useful for exploring in a more meticulous way the domain of interest and for observing the potential zone of convergence of the chains. 
The MH algorithm, unlike the Bayes Factor, explores the domain of interest and then it indicates the zone of higher probability, at MCMC convergence.

\subsection{ Posterior GPR }
\label{sec:disc1}
In Sec. \ref{sec:Results_it55} a detailed description of the results obtained with the MH algorithm is presented. The posterior plotted on Fig. \ref{fig:55_iter_GPR_histogram_joint} tends to concentrate close to $m_g=0$ with decreasing steps for larger $m_g$ up to $m_g < 0.15 \times 10^{-23} \; eV c^{-2}$. 
In contrast with the test case presented in Appendix \ref{sec:Results_it5}, the algorithm shows no detection for $m_g \neq 0$, at the GPR approximation of $\chi^2$.
For sake of completeness, in Appendix \ref{sec:Results_it5} we provide a test case for which the MH algorithm gives a detection as outcome of the simulations, both in terms of Bayes Factor and in terms of posterior density. If we had obtained a \af{positive} detection \af{(with an estimated value as mean of the posterior)}, it would have been similar to what appears in Appendix \ref{sec:Results_it5}. Since the outcome in Sec. \ref{sec:Results_it55} is totally different\om{---one does not have a Gaussian-like posterior centered on a positive value---}
we conclude \om{that we do not have a positive detection}
\af{, meaning that we can't provide an estimated value for the mass of the graviton, but we can give a 99.7$\%$ limit as quantile of the deduced mass posterior, $0.98 \times 10^{-24} \; eV c^{-2}$}.



\subsection{ Posterior GPUR }
\label{sec:disc2}
Next to a MCMC with a nominal $\tilde{\chi}^2$, in Sec. \ref{GP_and_int_uncertainty} is explored the possibility of accounting for the error of approximation, within the framework of GPR. The GPUR result presented in Sec. \ref{sec:Results_it55} is consistent with the nominal case, and it shows the posterior with a slightly larger interval of masses. This is actually what it is expected, since the uncertainty of GPR in the present case is small, but not absent. On Fig. \ref{fig:55_iter_300Pert_GPR_histogram_comparison} it is easy to see how much the maximum value of $m_g$ is {shifted} towards larger $m_g$, passing from the MCMC with GPR to the MCMC on GPUEs.
In particular the average going from $0.26 \times 10^{-24} \; eVc^{-2}$ to $0.34 \times 10^{-24} \; eVc^{-2}$ whereas the maximum mass in the posterior going from $1.45 \times 10^{-24} \; eVc^{-2}$ to $2.53 \times 10^{-24} \; eVc^{-2}$. 
The strategy we propose in Sec. \ref{GP_and_int_uncertainty} and \ref{sec:GPR_LinInt_Comparison} relies on the assumption that if we compute the real values of $\chi^2(m_g)$, then we can {estimate $\chi^2$ values in} the zones of domain for which $\chi^2(m_g)$ is unknown, with an uncertainty based on $\chi^2$ values already computed.
In our specific case, the strategy looks like consistent. The outcome of the 300 MCMC runs on different GPUEs (see Eq. \eqref{single_pert_form2}, Sec. \ref{GP_and_int_uncertainty} and Sec. \ref{sec:IT55_results_perturbations} ), is similar with respect to the nominal GPR case with slight differences (see e.g. Table \ref{tab:GPR_300iterations_comparison_55it} and Fig. \ref{fig:55_iter_300Pert_GPR_histogram_comparison}). 
\af{As previously indicated in Sect. \ref{sec:disc1}, again the GPUR posterior is not similar to the one obtained with a positive detection (See Appendix \ref{sec:Results_it5}). We are however able to provide limits for the mass.}
The upper bound for GPUR we would provide at $99.7\%$ CL is $m_g \leq 1.01 \times 10^{-24} \; eV c^{-2}$. \af{This represents an improvement of about 1 order of magnitude from the previous estimations in terms of CL, and constitutes the GPUR CL.} 

{As stressed in Table \ref{tab:GPR_300iterations_comparison_55it}, from a very large uniform prior between $0$ and $3.62 \times 10^{-23} \; eVc^{-2}$, the MCMC algorithm indicates a posterior between $0$ and a maximum of \vm{$2.53 \times 10^{-24} \; eVc^{-2}$}, inducing a significant improvement also  on the possible maximum value for the mass of the graviton.}
\subsection{ Laplace prior }
\label{sec:disc3}
\begin{figure}[!ht]
    \centering
    \includegraphics[width=.99\linewidth]{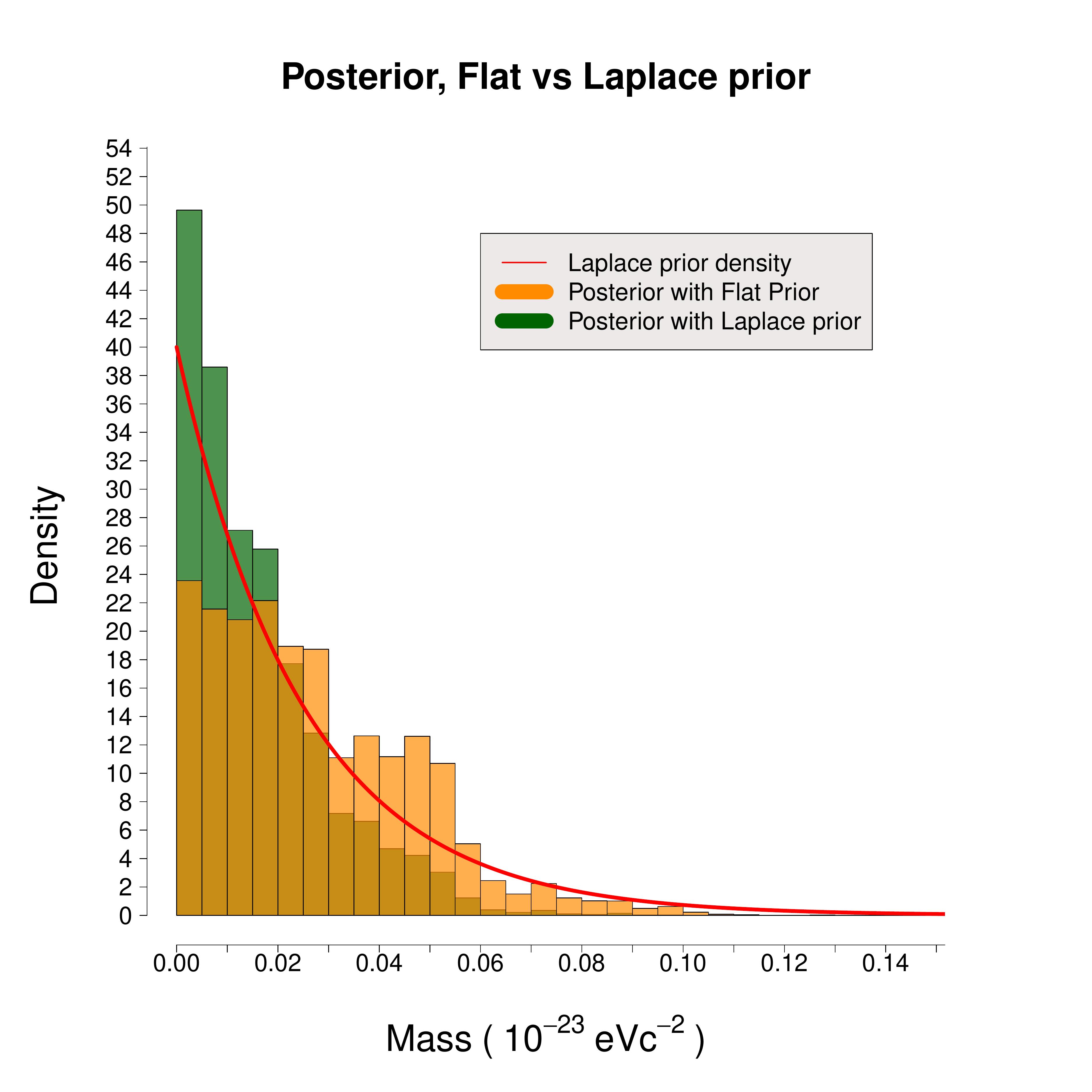}
    \caption{Densities for the posterior probability distributions obtained from GPR with a uniform (flat) prior (in orange) and with a half-Laplace prior (dark green). In red, the shape of the half-Laplace prior used. The brownish area is the overlaid zone of the two posteriors}
    \label{fig:55_iter_Flat_Laplace_priors_comparison}
\end{figure}

The \om{absence of} \af{positive} detection can be interpreted in two ways: either because the data employed are not sensitive enough, or because GRT is sufficient to explain the data. In order to discriminate between the former and the latter, we made a further validation test, changing the prior density for $m_g$ in the MH algorithm. We run the MH algorithm again, with the same GPR, but using a half-Laplace prior (red line on Fig. \ref{fig:55_iter_Flat_Laplace_priors_comparison}) \vm{instead of using a uniform prior}. 
On Fig. \ref{fig:55_iter_Flat_Laplace_priors_comparison} we see a comparison between the posterior obtained with a half-Laplace prior and the posterior with uniform prior. 
\vm{The half-Laplace prior used is chosen such that the zone of higher probability of the half-Laplace density shares the same domain of the posterior obtained with the uniform prior. In this way, the chains obtained with half-Laplace prior is a second step in the overall work, and it is not totally independent from the posterior with the uniform prior. \om{With the two posteriors,}
we may see whether the new posterior (green on Fig. \ref{fig:55_iter_Flat_Laplace_priors_comparison}) resembles the old one (orange on Fig. \ref{fig:55_iter_Flat_Laplace_priors_comparison}) or not: \om{we found that} it is not the case. The uniform (flat) prior let us provide an upper bound constraint on $m_g$, biasing as little as possible the algorithm outcome. On the other hand the half-Laplace prior has a different goal: it discriminates
\om{whether the data is not sensitive enough or GRT is sufficient to explain the data.}}
The MH algorithm does not provide the same outcome \vm{with the two different priors}. By mean of the half-Laplace prior, we are giving preference to the GRT\om{. Hence, because} the posterior tends to remain at the same density as the initial half-Laplacian, \om{it} means that GRT is sufficient to explain the data set. 
\af{In the interval of the GPUR CL obtained with the flat prior}, we can thus conclude that the information contained in the data set (and using this methodology) is not strong enough to move the narrow half-Laplace prior density towards a more flat density\om{---since the MH algorithm is choosing to remain stuck on GRT---}meaning that GRT is enough to explain completely the observational data \af{within this interval}. 
We conclude, then, that, \af{in the GPUR CL interval},  we do not have enough signal within INPOP to detect graviton mass \af{smaller than the threshold of $0.1 \times 10^{-23} \; eVc^{-2}$.} 

We refer to Sec. \ref{sec:Laplace_prior_5_chains} for a validation of the MCMC convergence in the case of half-Laplace prior.

\subsection{ Comparison with Bernus }
\label{sec:disc4}

\begin{figure}[!ht]
    \centering
    \includegraphics[width=.99\linewidth]{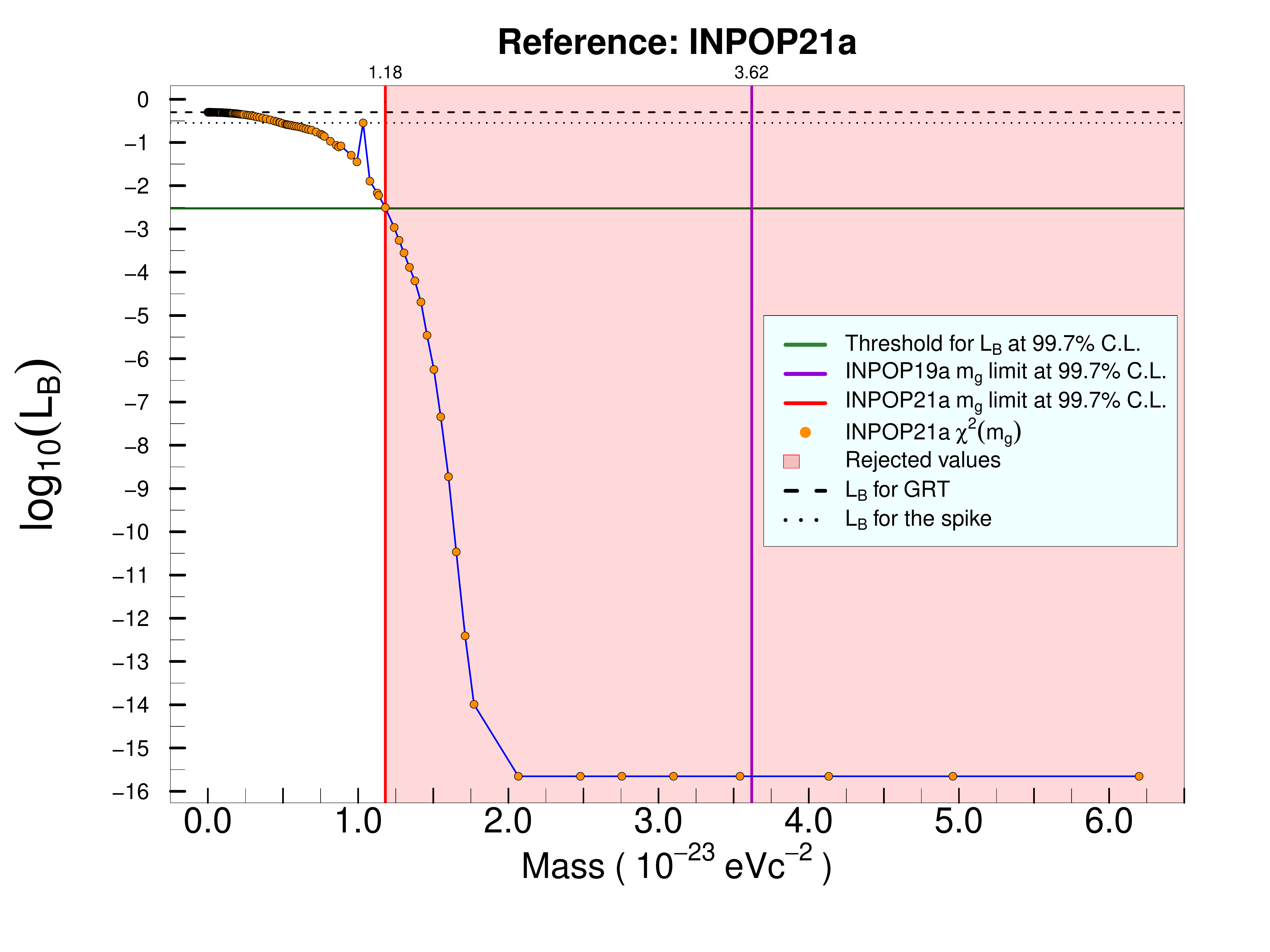}
    \caption{Values of the likelihood $L_B$ as computed in \cite{Bernus2020}. The orange dots represent the the $\log_{10}(L_B)$ values computed with INPOP21a. The green horizontal line is a threshold value for $L_B$ at $99.7 \%$ with the same criteria used in \cite{Bernus2020}. The vertical red line is the $m_g$ limit at $99.7 \%$ C.L. with the same criteria adopted in \cite{Bernus2020} but using INPOP21a. The violet vertical line is the  $m_g$ limit at $99.7 \%$ C.L. proposed in \cite{Bernus2020} using INPOP19a. The shaded red zone is the zone of rejected value we would obtain with INPOP21a at $99.7 \%$ C.L. . {The horizontal dashed line represent the $L_B$ value for GRT, whereas the horizontal dotted line represents the $L_B$ value for the spike around $m_g=1.03 \times 10^{-23} \; eVc^{-2}$.  } }
    \label{fig:Likelihood_Bernus_Comparison_v2}
\end{figure}

Bernus et al. in \cite{bernus2019, Bernus2020} have given upper bounds for the mass of the graviton obtained with INPOP17b and INPOP19a. In our work are presented a generalization and an improvement of the results given in \cite{bernus2019, Bernus2020}, using INPOP21a. 
By using a more general semi-Bayesian approach, we are showing that GRT is enough for explaining the data, and the massive graviton is not inducing any improvement in the planetary model. %
In order to compare with previous works, we can give an upper limit of the posterior.
In particular in \cite{bernus2019, Bernus2020} the upper bound for the mass with a $99.7 \%$ confidence level is $m_g \leq 3.62 \times 10^{-23} \;  eV c^{-2}$. 
In order to proceed with a comparison, we take the $99.7 \%$ confidence level on GPUR (see Sec. \ref{sec:IT55_results_perturbations} and Table \ref{tab:GPR_300iterations_comparison_55it}) which corresponds to a mass of $m_g \leq 1.01 \times 10^{-24} \;  eV c^{-2} $.
The result is an improvement by 1 order of magnitude in comparison with Bernus et al. \cite{Bernus2020}. It is however interesting to understand if this result is due to the Planetary Ephemerides improvement, or due to the change of methodology.
\om{Hence, we} used the formalism proposed by Bernus et al. in \cite{Bernus2020} with INPOP19a, but using INPOP21a $\chi^2$ values \om{instead}. 
Doing so we can understand whether the improvement \vm{of the constraint} is due to the MCMC method itself, or to the INPOP improvement. \cite{Bernus2020}  computed a "likelihood" interpreted as the probability of a tested theory to be likely. We refer to such a likelihood as $L_B$. The results obtained using INPOP21a are shown on Fig. \ref{fig:Likelihood_Bernus_Comparison_v2}. 
On this Figure, one can see that the INPOP21a $L_B$ value is improved by a factor 3, going from $3.62 \times 10^{-23} \; eVc^{-2}$ to $1.18 \times 10^{-23} \; eVc^{-2}$ for a C.L. at $99.7  \% $. A spike is present on Fig. \ref{fig:Likelihood_Bernus_Comparison_v2} for $m_g=1.03 \times 10^{-23} \;  eV c^{-2} $: this is due to a local minimum in the $\chi^2$ function. But let us stress that, even at this local minimum one has the likelihood $L_B$ (at $m_g = 1.03 \times 10^{-23} \;  eV c^{-2}$) that is still smaller than the likelihood $L_B$ obtained in GRT. Thus, GRT is still the most favourite guess between the two possibilities. The local minimum is also present in the function $m_g \longmapsto \tilde{\chi}^2(m_g)$ used in the MH algorithm: the MCMC process takes into account this local minimum and overcomes it going towards the global minimum at about $m_g \simeq 0$. For further details on the computation of $L_B$ see \cite{Bernus2020}. 
We can conclude that both the new model, and the new observations introduced with INPOP21a induce a factor 3 improvement relative to INPOP19a \af{Bernus determination} on $m_g$. The MH algorithm implemented here goes even further and explores a zone close to $m_g=0$.

\subsection{Comparison with {the LIGO-Virgo-KAGRA collaboration}}
\label{sec:disc5}

The {LIGO-Virgo-KAGRA} collaboration presents in \cite{LIGOVirgo_testsGRT_12_2021} an updated bound of the mass of the graviton $m_g$ at $90 \%$ credibility, that is $m_g \leq 1.27 \times 10^{-23} eV c^{-2}$. The posterior then obtained from the MCMC in our work seems to improve roughly by 1 order of magnitude this upper bound.
{It is important however to stress that the two studies are perfectly complementary because they}
focus on different aspects of the massive gravity phenomenology (radiative versus orbital), and use totally different observations (gravitational waves versus astrometry in the Solar System).

\section{Conclusion}

{We have presented our work on the use of MCMC algorithm and INPOP in order to get an improvement of the detection limit of the mass of the graviton $m_g$ using the Solar System dynamics as arena. 
A key strength of the present study is to include the $m_g$ contribution in terms of accelerations and light times computation within the full dynamics of the Solar System. Moreover, by considering a semi-Bayesian approach, and using MCMC and MH algorithm, we avoid correlations between $m_g$ and other INPOP astronomical parameters.
We used the GPR to obtain an approximation of the $\chi^2$ ready to use within the MH algorithm and to asses the uncertainty of approximation afterwards. Beside the MCMC approach, the Bayes Factor has been computed as well with unapproximated $\chi^2$ values.
From the posterior obtained, we can give an upper bound of $m_g \leq 1.01 \times 10^{-24} \; eVc^{-2}$ at $99.7 \%$ CL (resp. $\lambda_g \geq 122.48 \times 10^{13} \; km$), including approximation uncertainty, and we had shown with a change of the prior (from flat to  half-Laplace) that no significant information is detectable in planetary ephemerides for masses smaller than this limit. }

Observations from the BepiColombo mission in the coming years will provide new data to improve the overall INPOP fit and also the constraints on the INPOP parameters.

\section{Acknowledgments}
{VM was funded by CNES (French Space Agency) and UCA EUR Spectrum doctoral fellowship. This work was supported by the French government, through the UCAJEDI Investments in the Future project managed by the National Research Agency (ANR) under reference number ANR-15-IDEX-01. The authors are grateful to the OPAL infrastructure and the Université Côte d’Azur Center for High-Performance Computing for providing resources and support. VM and AF thank  A. Chalumeau, C. Twardzik, S. Babak and L. Bigot for their useful inputs and discussions.}

\bibliography{Bibliography_natbib}

\appendix

\section{Notations}

\renewcommand{\thefigure}{A\arabic{figure}}
\setcounter{figure}{0}
\renewcommand{\thetable}{A\arabic{table}}
\setcounter{table}{0}

In this section we gather the different notations used in Sect. \ref{sec:th}. The more important symbols we used are summarized in Table \ref{tab:Table_notation}. 

\begin{table}[] \scriptsize  
\centering
\caption{Summary of the more relevant notation used.}
\begin{tabular}{|p{2cm}|p{5cm}|}  
 \hline
 \centering Notation & Description \\ [0.5ex] 
 \hline\hline 
\centering $\chi^2$ & The function of the normalized $\chi$-squared: for a given value $m_g$ of the graviton mass  $\chi^2(m_g)$ is computed fitting all the remaining astronomical parameters of INPOP with the observations, using the INPOP numerical integration and least squares processes. \\ 
 \hline
 \centering $\tilde{\chi}^2$ & The approximation of the function  $\chi^2$ used in the actual implementation of the MH algorithm to obtain the posterior. As approximation we used the GPR starting from a set of points $S$. \\ 
 \hline

 \centering $\breve{\chi}^2$ & The symbol indicates a GPUE. In other words it is one \emph{realization} of the GPR: it is a perturbation of the nominal solution to the interpolation provided by the GPR. The construction is based on the set $\breve{S}^{M}$ in Eq. \eqref{single_pert_form1} and explained in Section \ref{GP_and_int_uncertainty}. \\ 
 \hline


  \centering  $\chi^2(m_g)$ & It is the value of the function $\chi^2$ evaluated in the point $m_g$.  \\ 
 
 \hline

 \centering  $\tilde{\chi}^2(m_g)$ & It is the value of the function $\tilde{\chi}^2$ evaluated in the point $m_g$.  \\
 \hline

  \centering $\breve{\chi}^2(m_g)$ & It is the value of the function $\breve{\chi}^2$ evaluated in the point $m_g$. Generally speaking $\breve{\chi}^2(m_g) \neq \tilde{\chi}^2(m_g) $ .\\
\hline

  \centering $b_l$ & Lower bound for the mass $m_g$. This value is the value of the mass for which if $m^*_g < b_l$ is proposed during the MH algorithm then it is automatically not accepted.\\
\hline

  \centering $b_u$ & Upper bound for the mass $m_g$. This value is the value of the mass for which if $m^*_g > b_u$ is proposed during the MH algorithm then it is automatically not accepted.\\
\hline

  \centering $m_g^i$ & We indicate with $m_g^i$ the single element of the mesh chosen to built the perturbation of the GPR, as described in Sec. \ref{GP_and_int_uncertainty}. \\
\hline

  \centering $\overrightarrow{m}_g$ & Vector of masses for which we compute $\chi^2$ without approximation (using the full INPOP). \\
\hline

  \centering $ m_g \longmapsto \tilde{\sigma}_i(m_g) $ & Function of $1\sigma$ interpolation uncertainty provided by the GPR. \\
\hline

  \centering $ \chi^2(\overrightarrow{m}_g) $ & Vector of components $\chi^2(m) \; \forall m \in \overrightarrow{m}_g$. \\
\hline

\end{tabular}
\label{tab:Table_notation}
\end{table}

\section{Validation of the algorithm} \label{sec:Results_it5} 

\renewcommand{\thefigure}{B\arabic{figure}}
\setcounter{figure}{0}
\renewcommand{\thetable}{B\arabic{table}}
\setcounter{table}{0}

\begin{figure}[!ht]
\centering
  \includegraphics[width=.99\linewidth]{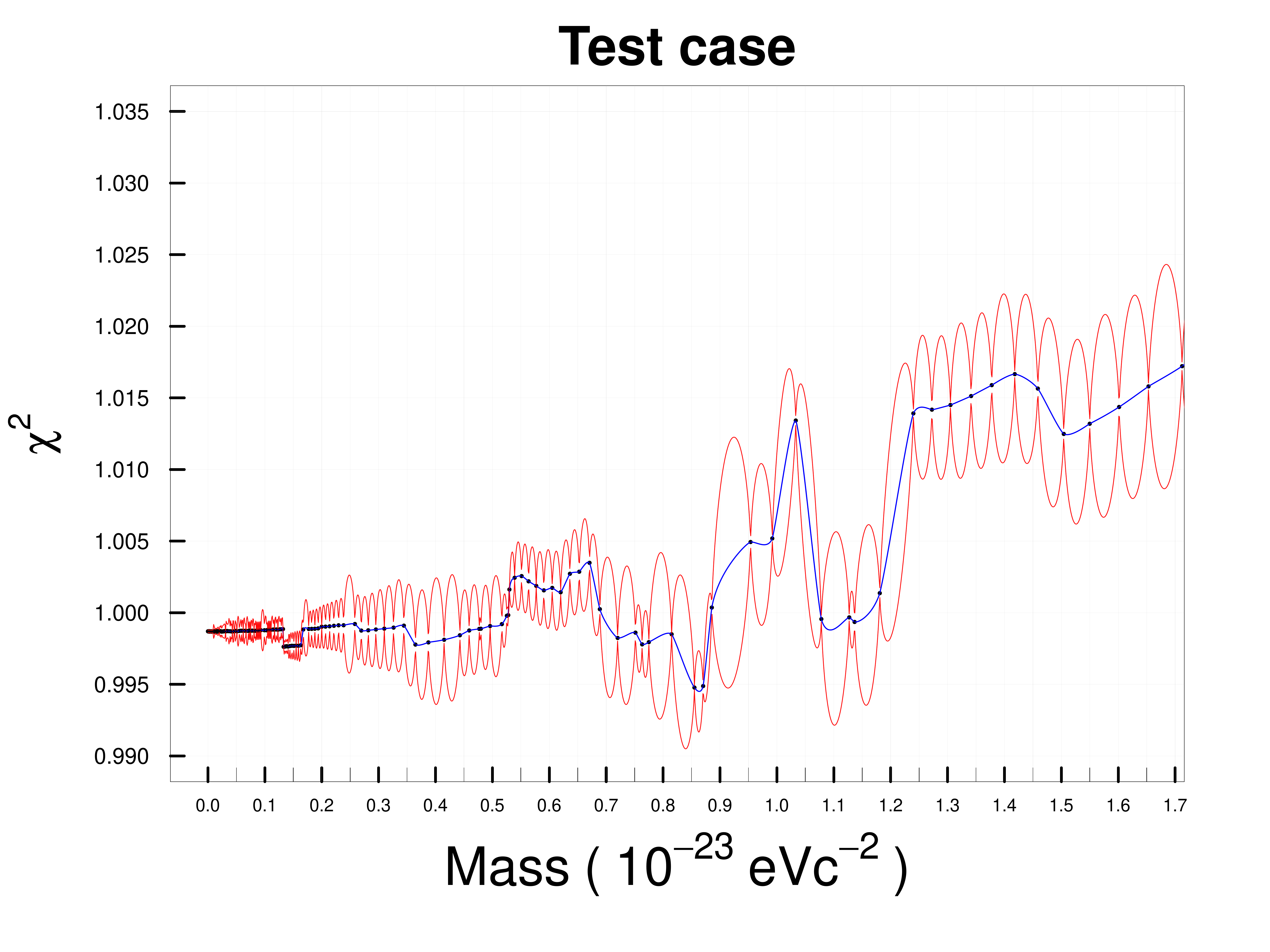}
   \caption{Plot of the function $ m_g \longmapsto \tilde{\chi}^2(m_g)$. The dots correspond to $\chi^2$ value estimated with INPOP and not extrapolated. the red lines correspond to the $2-\sigma$ uncertainties provided by the GPR. The set $S$ is the one used for the test case.}
    \label{fig:GPR_only_plot_IT5_v3}
\end{figure}

In order to validate the MCMC implementation, we run the algorithm on an example of INPOP simulations presenting a $\chi^{2}$ minimum for a value of $m_g$ different from 0 as illustrated on Fig. \ref{fig:GPR_only_plot_IT5_v3}. On this Fig., the minimum of the $\chi^{2}$ is reached for $m_g \simeq 0.86 \times 10^{-23} \; eV c^{-2}$. This case, called in the following {\it{test case}}, has been obtained during the iterative procedure of the INPOP \af{construction but before the full convergence of the adjustment. It is  used} here for sake of demonstration of the algorithm capabilities for detecting possible local or global minima.
The procedure followed is analogous to what we already proposed in Sec. \ref{sec:BF_results_it55} and Sec. \ref{sec:IT_55_GPR_MCMC}, starting from a different set $S$. The difference between the set $S$ of the real case ($\chi^2$ after INPOP fit adjustment) and the set $S$ used in the test case is in the values of INPOP $\chi^2$ associated to the masses, thus in the GPR interpolation we obtain.

\subsection{Bayes Factor} \label{sec:BF_results_it5}
\begin{figure}
    \centering 
    \includegraphics[width=.99\linewidth]{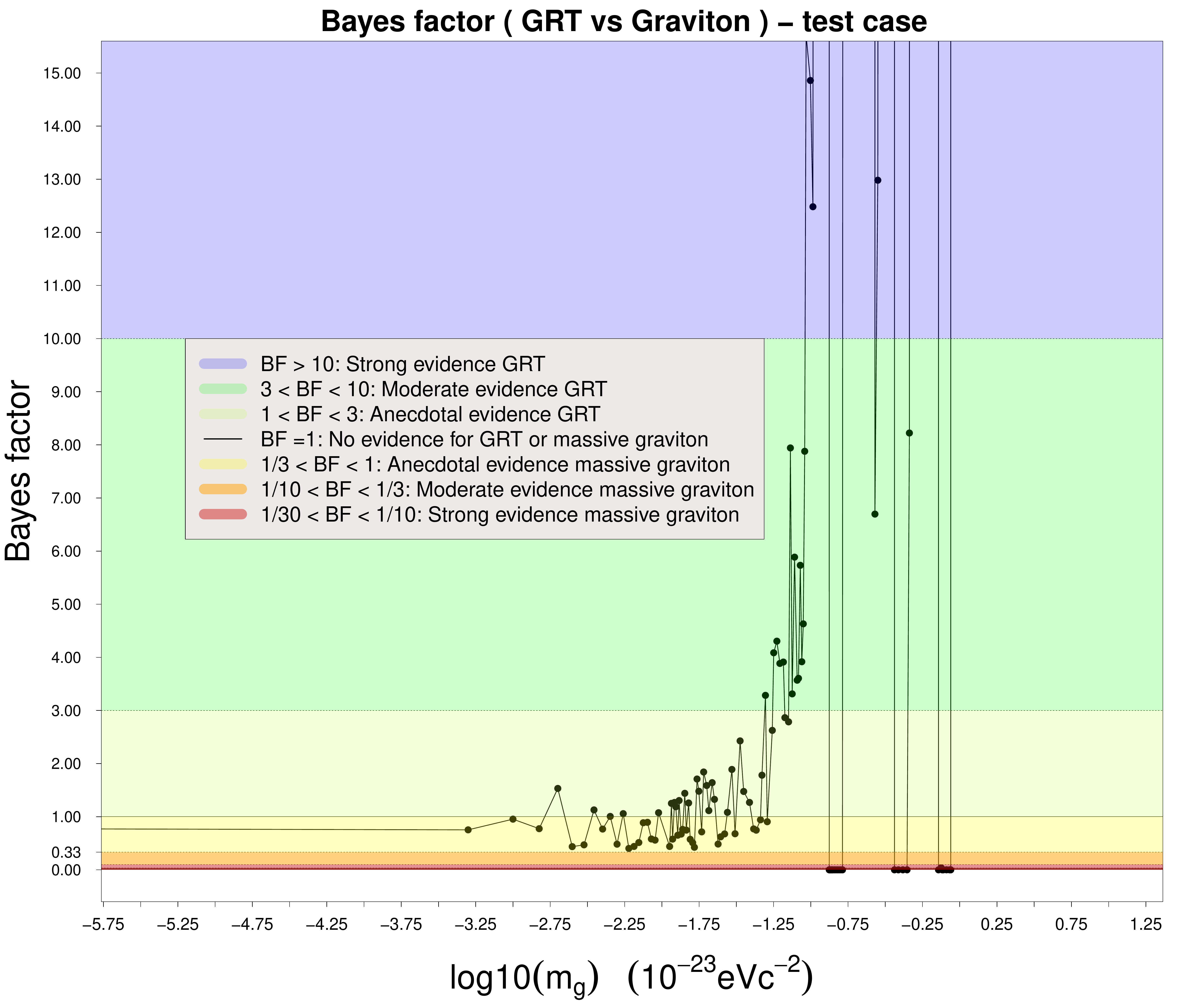}
    \caption{Value of the Bayes Factor as function of $m_g$ for the test case. Each point corresponds to one INPOP run.}
    \label{fig:BayesFactor5}
\end{figure}

As we see on Fig. \ref{fig:BayesFactor5}, the results obtained for different values of $m_g$ in the test case are an interesting example. It appears clearly on \ref{fig:BayesFactor5} that a minimum of $\chi^2$ is reached for some non-zero values of $m_g$. This means that for some values of $m_g$ the $\chi^2$ estimated in the test case is quite smaller that the $\chi^2$ obtained with the corresponding GRT value. These comments do not imply that the graviton gives better results than GRT, but just that for the test case, 
it gives an example of how{, when a graviton mass gives better residuals than GRT, it can be detected from the Bayes Factor.}
In accordance to the notation presented in Sec. \ref{sec:BF_theory}, as Model 1 we use GRT,
whereas as Model 2 we employ a massive graviton with $m_g$ assuming the value of one component of the vector $\overrightarrow{m_g}$. Model 1 corresponds to GRT,  and Model 2 corresponds to an alternate possible value of $m_g \neq 0$.
The Bayes Factor for the test case is shown on Fig. \ref{fig:BayesFactor5}. On Fig. \ref{fig:BayesFactor5}, the value of the Bayes Factor tends to be \emph{below} $1$, and specifically below $\frac{1}{10}$ for values of the mass around $m_g= 0.86 \times 10^{-23}  \; eV c^{-2}$. 


{But first, let us stress that} that the Bayes Factor is a qualitative tool, that has not generally accepted interpretation (see, e.g. \cite{GillJeff2008}) and it is in any case \emph{just} a comparison among two specific values (Model 1 vs Model 2).
{Nevertheless, one could think that a model with $m_g \sim 0.86 \times 10^{-23}  \; eV c^{-2}$ has more evidence than GRT. But let us stress that the test case considered in Fig. \ref{fig:BayesFactor5} was obtained before full convergence, and that this specific minimum of the $\chi^2$ value at $m_g \sim 0.86 \times 10^{-23}  \; eV c^{-2}$ disappears after full convergence---as mentioned at the beginning of Sec. \ref{sec:Results_it5}.}
Hence, this test case can be seen as a good example of a fake detection since the $\chi^2$  evaluated in this value of the graviton mass is indeed smaller than the GRT $\chi^2$ value. 
In the case of a global minimum of $\chi^2$ is reached for a given value of $m_g \neq 0$, the Bayes factor gives a positive detection. 
%
%
The difference between Fig. \ref{fig:BayesFactor5} (test case) and Fig. \ref{fig:BayesFactor55} (after INPOP fit adjustment) is in the  $\chi^2(m_g)$ values used. This means that even at a qualitative level, the interpretation of the results changes substantially.

\subsection{GPR and MCMC} \label{sec:IT5_Results_MCMC}
\begin{figure}
    \centering
    \includegraphics[width=.99\linewidth]{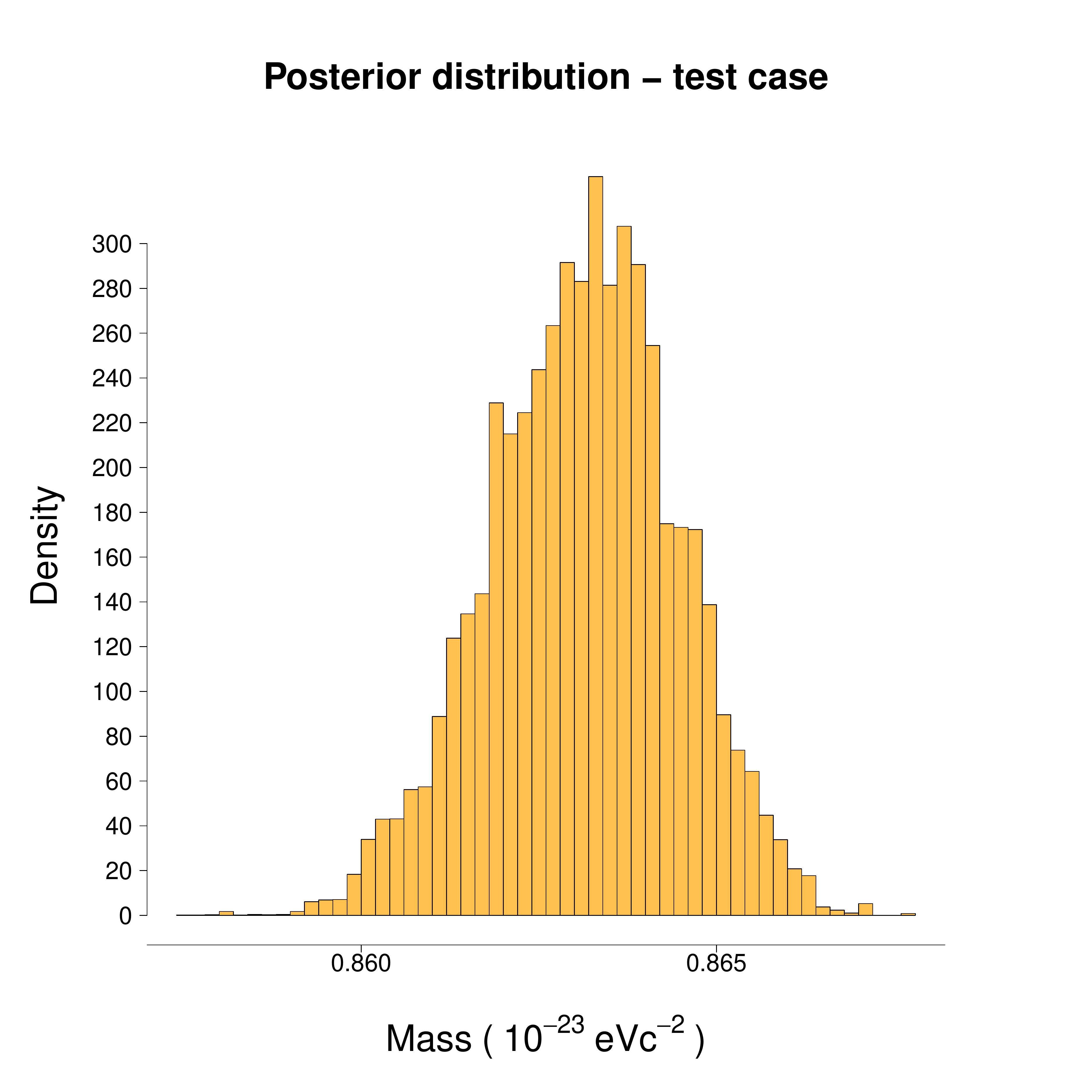}
    \caption{Final posterior distribution for the test case. This posterior density is what we expect in case of detection for an $m_g \simeq 0.86 \times 10^{-23} \; eVc^{-2}$}
    \label{fig:5_iter_GPR_histogram_joint}
\end{figure}

We present here the simulations of the MCMC computed for the test case, meaning with a global minimum of $\chi^2$ reached for a given value of $m_g \neq 0$.
The shape of $m_g \longmapsto \tilde{\chi}^2(m_g)$ is given on Fig. \ref{fig:GPR_only_plot_IT5_v3}. In this case the global minimum of $\tilde{\chi}^2$ is obtained for a graviton mass of about $0.86 \times 10^{-23}  \; eV c^{-2}$.
In particular we expect the chains output to have a distribution of values around this point.  On Fig. \ref{fig:5_iter_GPR_histogram_joint} is presented the posterior distribution from the combined 5 final posteriors (similary to Sec. \ref{sec:IT_55_GPR_MCMC}). We can easily see how the chain is well distributed around a central value, that is $0.863 \times 10^{-23} eVc^{-2}$.
The shape of the posterior is a bell shape, very close to what a normal distribution could be. This is the result that we expected to obtain since the $\tilde{\chi}^2$ has a global minimum quite narrow. Considering proper starting points for the chains and proper $\sigma_1$, the standard deviation of the normal distribution used for proposing new masses in the MH algorithm (see function $q$ in Eq. \eqref{eq:GaussianProposal}), we obtained convergence by use of $10^4$ steps in the MH algorithm. 
As it is pointed out from Fig. \ref{fig:5_iter_GPR_histogram_joint}, using the specific $\tilde{\chi}^2$, we would have had an evidence toward a massive graviton of $(0.8633 \pm 0.0016 ) \times 10^{-23} \; eVc^{-2} $, where $0.0016 \times 10^{-23} \; eVc^{-2}$ is the standard deviation of the density on Fig. \ref{fig:5_iter_GPR_histogram_joint}. 
This result obtained with the test case shows that our MCMC algorithm is indeed pointing towards the good minimum of the $\chi^2$ distribution with appropriate posterior density distribution, after tuning what was necessary properly, and checking MCMC convergence criteria. 
Furthermore the Bayes factor seems also to be able to indicate a positive detection in the context of this false improvement of the $\chi^2$.

\begin{table}
\centering
\caption{\label{tab:GPR_5it} Summary of the outcome divided by chain. $\tilde{\chi}^2$ taken as GPR. Test case. The unit for $m^I_g$, $\langle m_g \rangle $ and $97.7 \%$ quantile is $10^{23} \; eVc^{-2}$ }
\begin{tabular}{| c || c | c | c | c | c | c | }
\hline
 Chain &  $m^I_g $ & $ \langle m_g \rangle $ & Acc. rate & $97.7 \% $ quantile  \\ [1.2ex]
\hline \hline 

1 & $0.882 $ &  $ 0.8633 $ & $53.6 \%$ &  $0.868 $  \\ \hline
2 & $0.827 $ &  $ 0.8633 $ & $53.1 \%$ &  $0.867 $  \\ \hline
3 & $0.856 $ &  $ 0.8633 $ & $53.6 \%$ &  $0.867 $  \\ \hline
4 & $0.847 $ &  $ 0.8633 $ & $53.4 \%$ &  $0.867 $  \\ \hline
5 & $0.886 $ &  $ 0.8633 $ & $53.6 \%$ &  $0.867 $  \\ \hline

\end{tabular}
\end{table}


\section{Detailed results of 5 MCMC chains}
\renewcommand{\thefigure}{C\arabic{figure}}
\setcounter{figure}{0}
\renewcommand{\thetable}{C\arabic{table}}
\setcounter{table}{0}

In this Appendix we show some technical details about the convergence of the MCMC presented in Sec. \ref{sec:Results_it55} and Sec. \ref{sec:Results_it5}. For both cases, we provide a more detailed analysis of the convergence diagnostic tools used. As already mentioned in Sec. \ref{sec:MCMC_diagnosis}, we run five chains, in parallel, with different starting points. The starting points have been chosen in order to speed up the MH algorithm convergence with a relatively small number of iterations. The standard deviation $\sigma_1$ for the proposal normal distribution (see Eq. \eqref{eq:GaussianProposal}) has been tuned accurately to have good outcomes in terms of acceptance rate of the chaiins and sparsity within the zone of high probability. For each one of the two cases (INPOP case after fit adjustment and test case) we will show the details of the five chains separately.

\subsection{INPOP case, after fit adjustment} \label{app:sec:results_it_55}
\begin{figure}[!ht]
    \centering
    \includegraphics[width=0.9\textwidth]{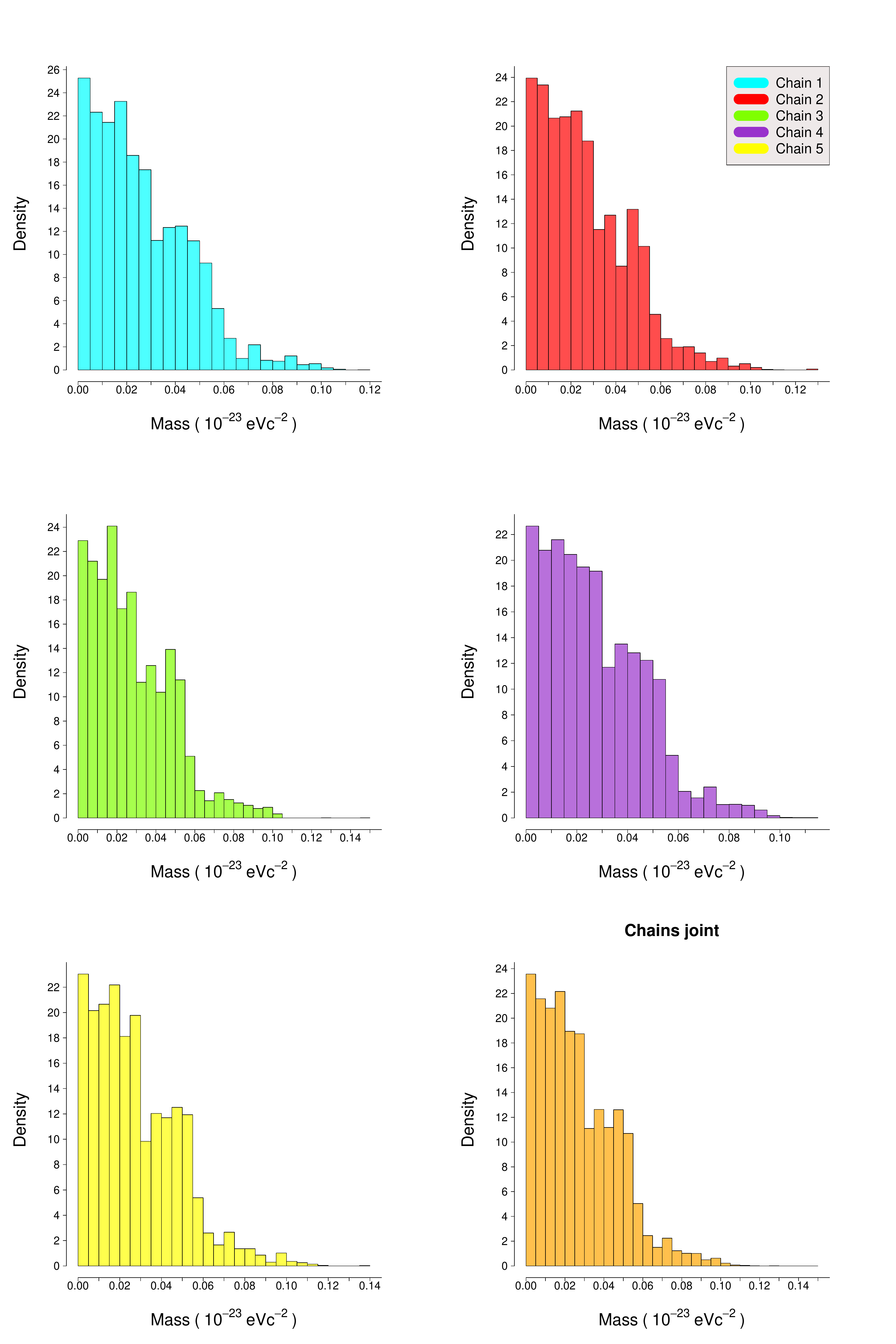}
    \caption{Density for the posterior probability distribution used as target probability in the MH algorithm for the five chains.}
    \label{fig:55_iter_GPR_histogram_5c}
\end{figure}

\begin{figure}[!ht]
    \centering
    \includegraphics[width=0.9\textwidth]{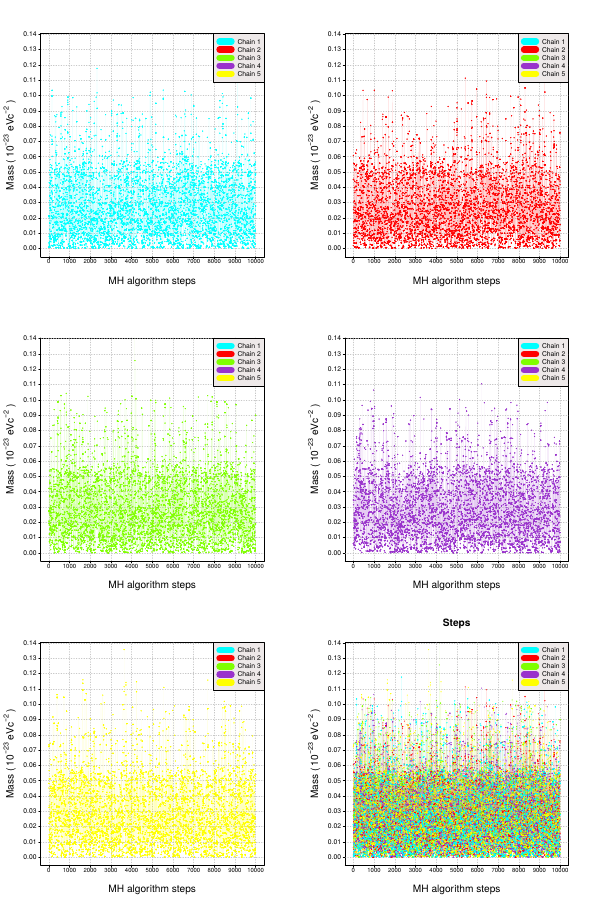}
    \caption{Traces of the five chains after having discarded the burn-in separeted by chain.}
    \label{fig:Sequence_of_the_chains_it55_GPR_single}
\end{figure}

\begin{figure}[!ht]
    \centering
    \includegraphics[width=0.9\textwidth]{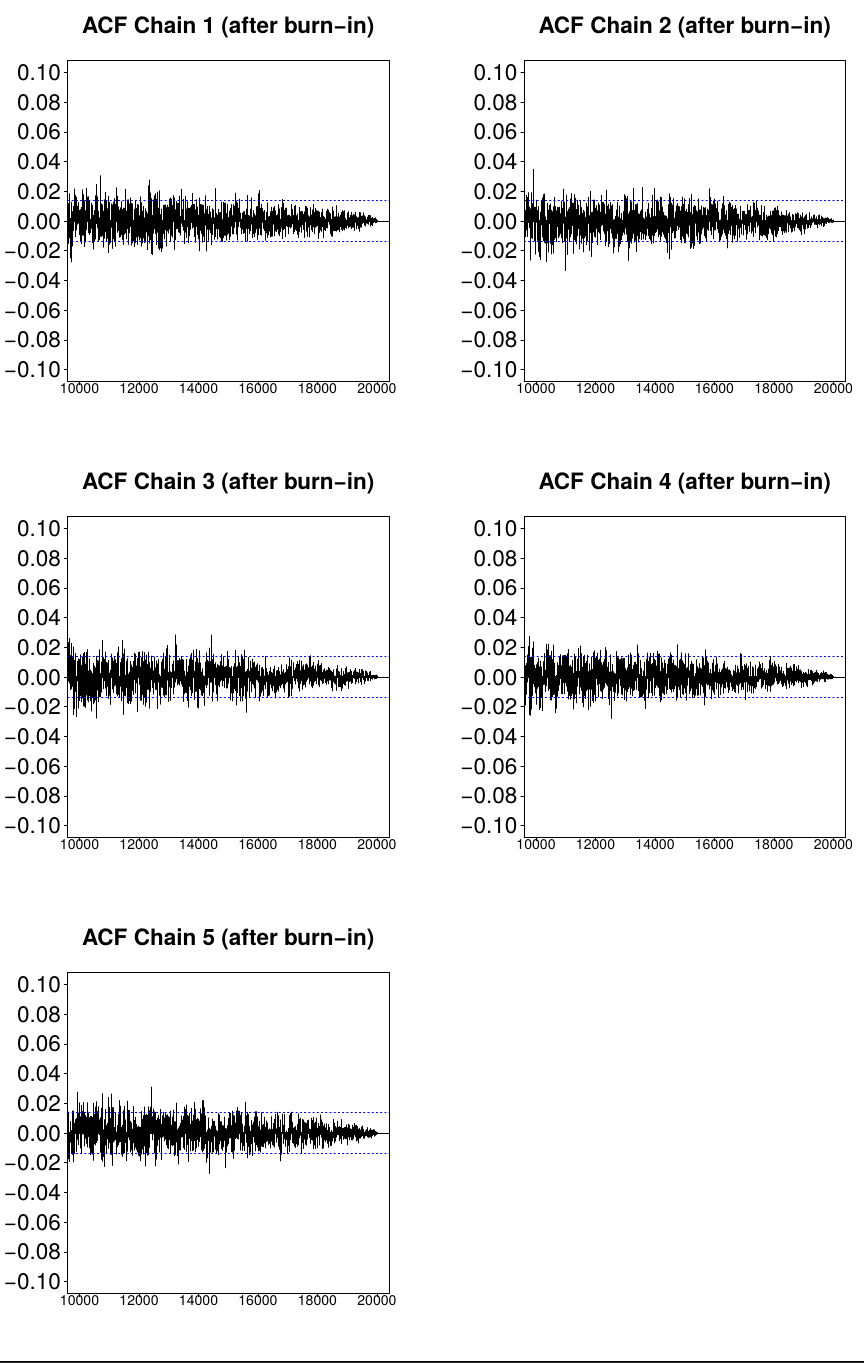}
    \caption{Autocorreletion function for the five chains run in parallel plus the density of the five chains joint. The plots are a zoom on the second half of each chain. The ACF function in these cases are always below $0.04$ (in absolute value). }
    \label{fig:ACF_plots_it55_GPR}
\end{figure}

In this section are presented the results obtained for the 5 different chains that constitute the final result presented in Sec. \ref{sec:BF_results_it55}. On Fig. \ref{fig:55_iter_GPR_histogram_5c} the posterior density of each chain is shown separately, except for the last plot (bottom right) that shows the density of the five chains merged, as already presented on Fig. \ref{fig:55_iter_GPR_histogram_joint}. 
As specified in Table \ref{tab:GPR_55it} (see Sec. \ref{sec:IT_55_GPR_MCMC}) the averages of the 5 posteriors are all about $\langle m_g \rangle \simeq 0.026-0.027 \times 10^{-23} \; eV c^{-2}$, varying from $0.0261 \times 10^{-23} \; eV c^{-2}$ to $0.0278 \times 10^{-23 \; eV c^{-2}}$. Similarly to the mean values, the $99.7 \%$ quantiles vary from $0.090 \times 10^{-23} \; eV c^{-2}$ to $0.098 \times 10^{-23} \; eV c^{-2}$. These figures, together with the shapes of the posterior densities shown on Fig. \ref{fig:55_iter_GPR_histogram_5c}, tell cleary that the 5 independently run chains are very similar among them. On Fig. \ref{fig:ACF_plots_it55_GPR} the plots of autocorrelation functions (ACF) are presented. The ACF is computed on the whole chains, from $0$ up to $3 \times 10^4$. On Fig. \ref{fig:ACF_plots_it55_GPR} each plot shows iterations after burn-in since it is the part we used to compute the posterior density of probability. The ACF  on this interval takes at most the value of $0.04$. Looking at Fig. \ref{fig:Sequence_of_the_chains_it55_GPR_single} we see the traces of the five chains used to obtain the final posterior on Fig. \ref{fig:55_iter_GPR_histogram_joint}. The traces show for each step of the MH algorithm (so each time a new value of the chain is decided) the outcome of the choice itself. The plots are quite sparse inside the zone for which we are interested in and this is a condition to have good convergence of the MCMC process. Finally, the acceptance rates of the chains are satisfactory: they span between $36 \%$ and $37 \%$. Let's recall that the Gelman-Rubin ratio is $1.000928$. In light of the results obtained we are confident that the chains converged.

\subsection{Validation case - test case}

\begin{figure}[!ht]
    \centering
    \includegraphics[width=0.9\textwidth]{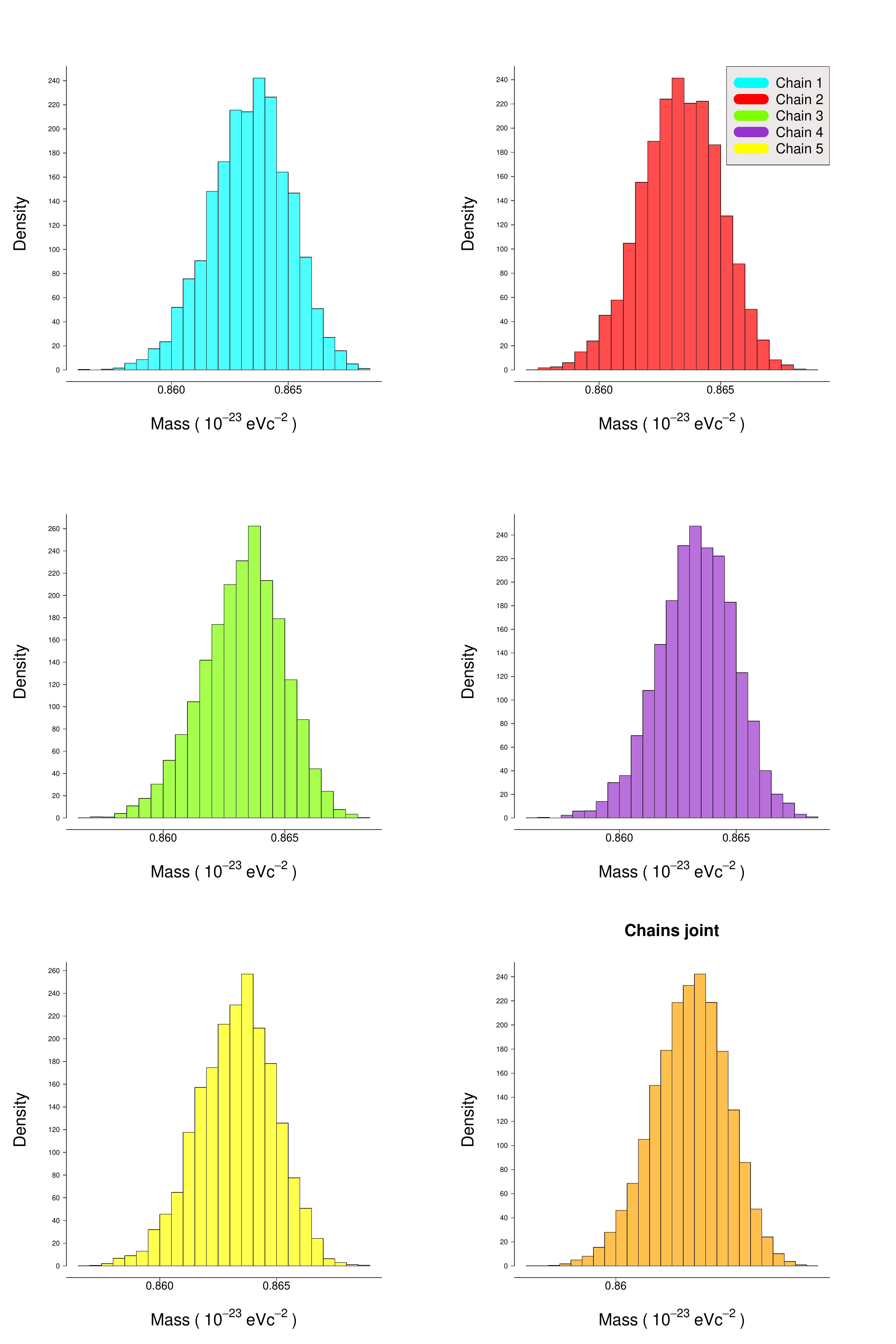}
    \caption{Posterior distributions of the five chains for the test case.}
    \label{fig:5_iter_GPR_histogram_5c}
\end{figure}

\begin{figure}[!ht]
    \centering
    \includegraphics[width=0.9\textwidth]{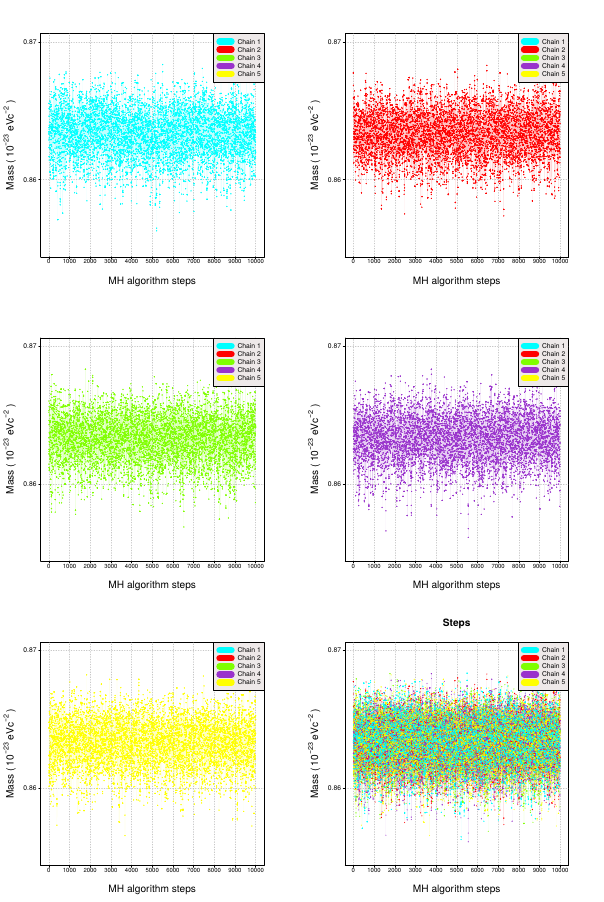}
    \caption{Traces of the five chains computed for the test case.}
    \label{fig:5_iter_GPR_sequence_of_the_chains_single}
\end{figure}

\begin{figure}[!ht]
    \centering
    \includegraphics[width=0.9\textwidth]{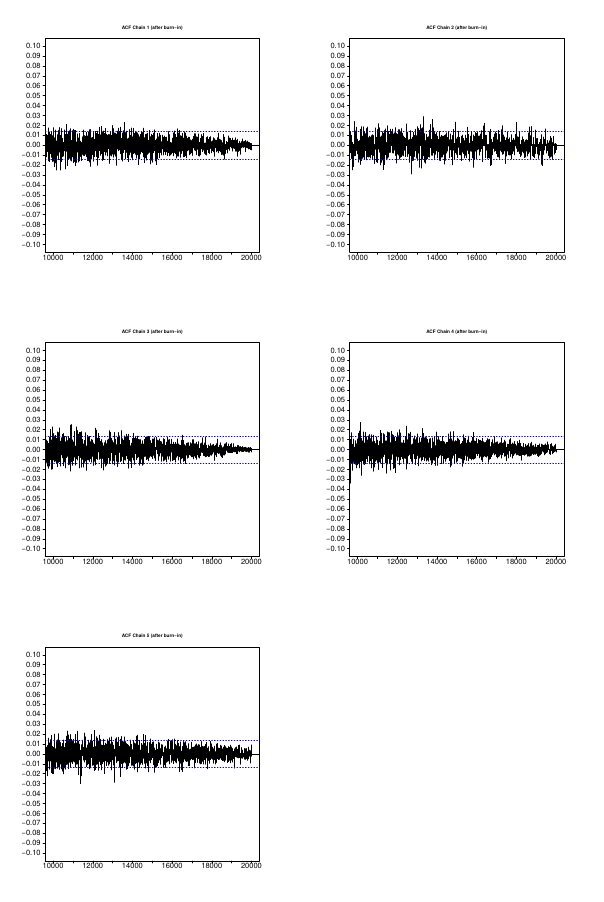}
    \caption{ACF plots of the five chains for the test case. In terms of absolute value, the ACF is always below 0.04}
    \label{fig:5_iter_GPR_ACF_plots}
\end{figure}

On Fig. \ref{fig:5_iter_GPR_histogram_5c} the posterior distributions of the five chains for the test case are presented. We see that they do not differ substantially in the shapes, and this is summarized indeed in Table \ref{tab:GPR_5it}. The averages in the five cases are always the same and the $99.7 \%$-quantile is between $0.867$ and $0.868$ for all the chains. The Gelman-Rubin factor $\hat{R} = 1.000151
$. On Fig. \ref{fig:5_iter_GPR_sequence_of_the_chains_single} the traces of the five chains are shown. We can see that the chains are quite sparse also in this case, and the zone they span is where we find the well of global minimum in the $\tilde{\chi}^2$.
As final analysis we show on Fig. \ref{fig:5_iter_GPR_ACF_plots} the autocorrelation functions (ACF) for the second half of each chain. We see that ACF plots behave well and autocorrelation is quite low. In order to obtain such a result is of essential importance to tune properly the number $\sigma_1$ since on it is based the speed of exploration inside the domain. Also in this case, we chose the initial points of each chain in order to speed up the convergence process of the MCMC, such that we could obtain it with less iterations. A properly tuned $\sigma_1$ is helpful both for ACF plots and for the acceptance rates of the chains. We recall from Table \ref{tab:GPR_5it} that the acceptance rates are in this case about of $53\%$. Also in the test case, we have reached MCMC convergence.

\subsection{Using a half-Laplace prior}\label{sec:Laplace_prior_5_chains}

We provide here the elements to assess the MCMC convergence in the case of a half-Laplace prior in the MH algorithm.

\begin{figure}[!ht]
    \centering
    \includegraphics[width=0.9\textwidth]{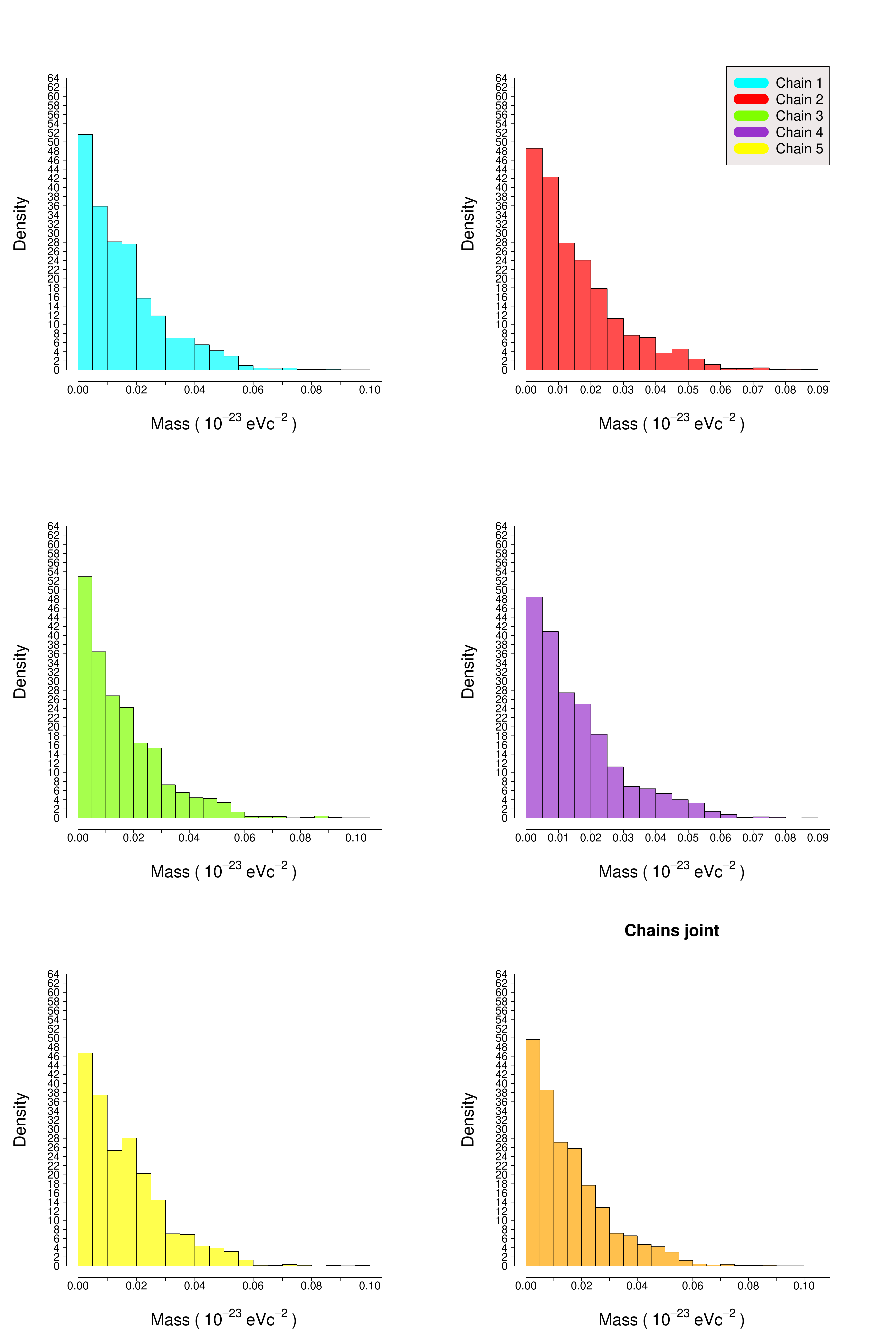}
    \caption{Posterior distributions of the five chains. In the MH algorithm the half-Laplace prior has been used.}
    \label{fig:Laplace_prior_GPR_histogram_5c}
\end{figure}

\begin{figure}[!ht]
    \centering
    \includegraphics[width=0.9\textwidth]{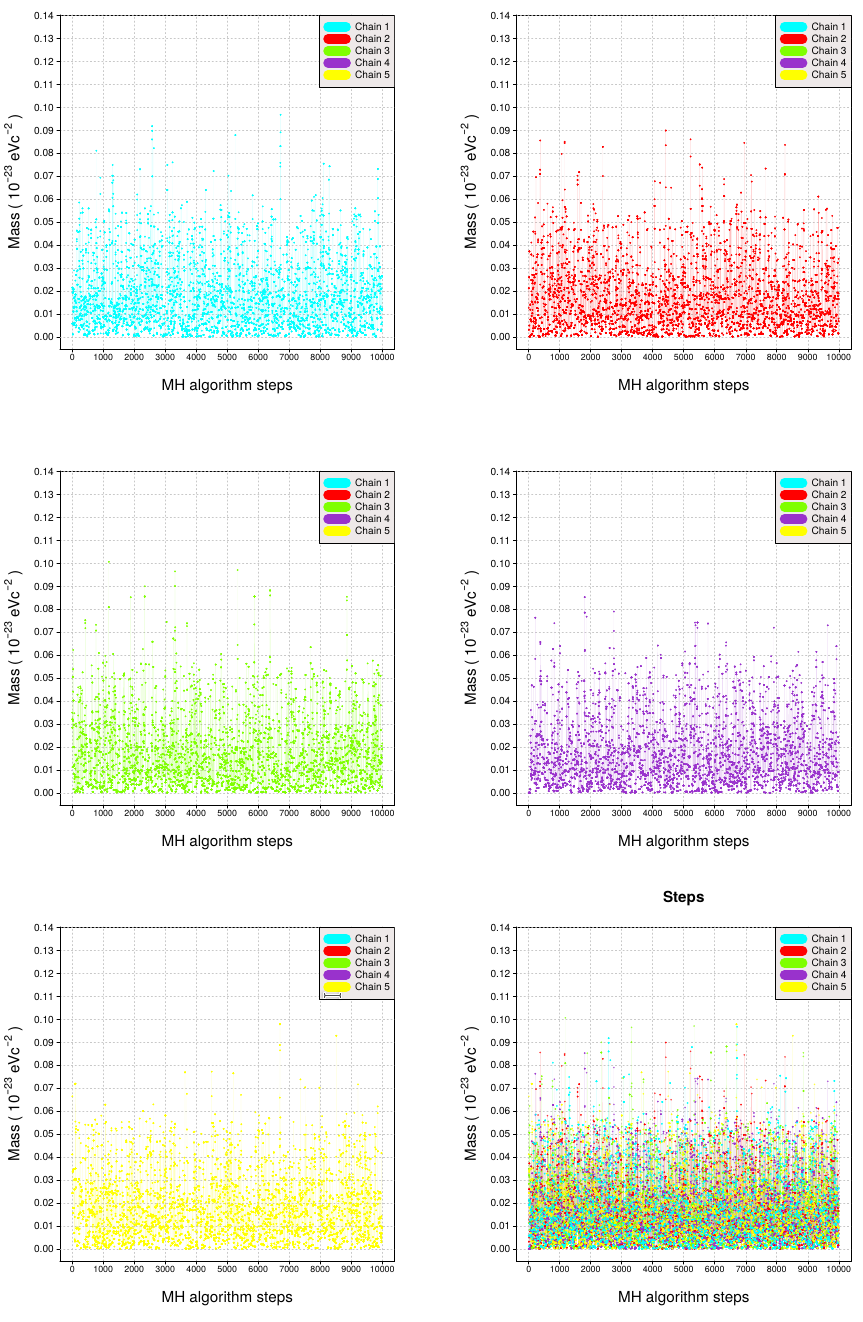}
    \caption{Traces of the five chains computed. In the MH algorithm the half-Laplace prior has been used.}
    \label{fig:Laplace_prior_GPR_sequence_of_the_chains_single}
\end{figure}

\begin{figure}[!ht]
    \centering
    \includegraphics[width=0.9\textwidth]{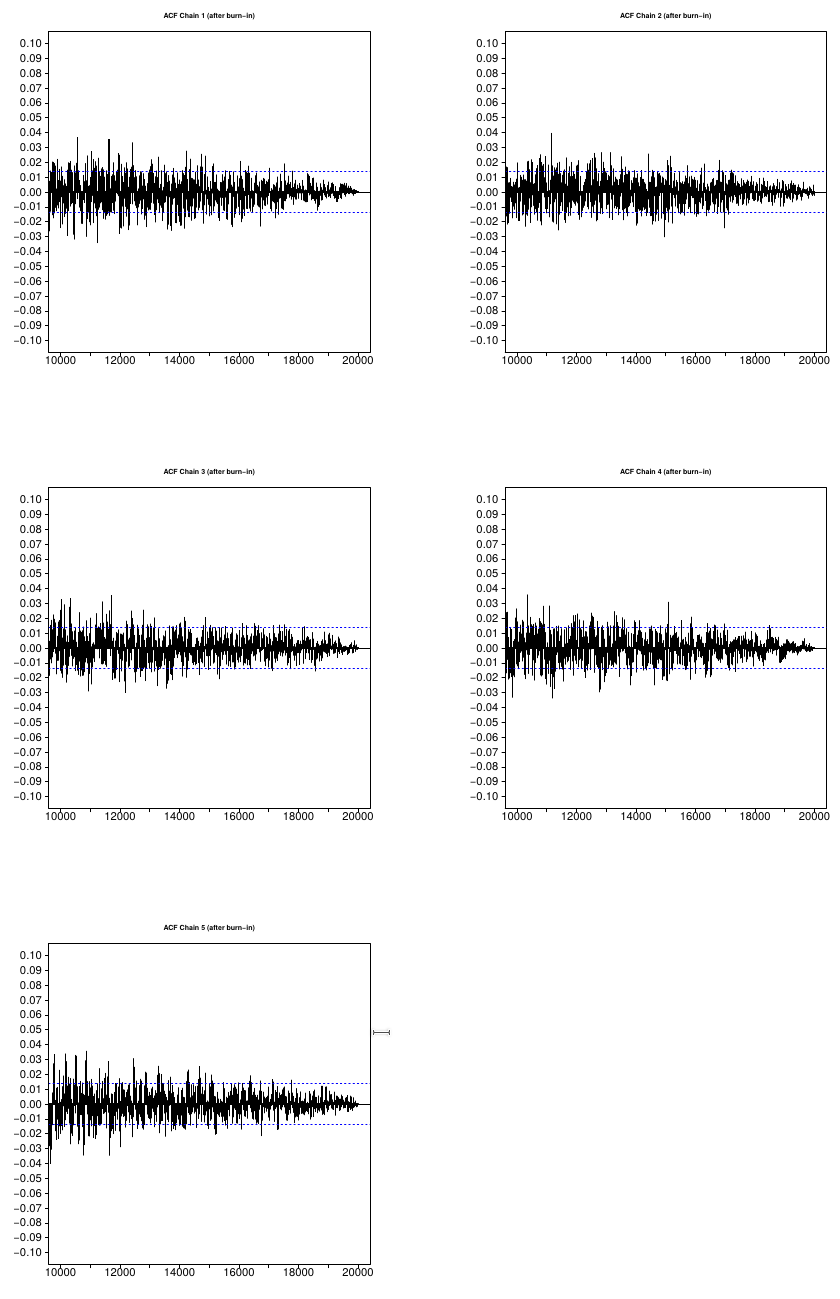}
    \caption{ACF plots of the five chains. In the MH algorithm the half-Laplace prior has been used.}
    \label{fig:Laplace_prior_GPR_ACF_plots}
\end{figure}

On Fig. \ref{fig:Laplace_prior_GPR_histogram_5c} the posterior distributions of the five chains for the test case are presented. 
We see that they do not differ substantially in the shapes, and this is summarized indeed in Table \ref{tab:posterior_with_Laplace}. The averages in the five cases are always the same and the $99.7 \%$-quantile is between $0.06 \times 10^{-23} \; eVc^{-2}$ and $0.08 \times 10^{-23} \; eVc^{-2}$ for all the chains. The Gelman-Rubin factor $\hat{R} = 1.000086$. On Fig. \ref{fig:Laplace_prior_GPR_sequence_of_the_chains_single} the traces of the five chains are shown. We can see that the chains are quite sparse.
As final analysis we show on Fig. \ref{fig:Laplace_prior_GPR_ACF_plots} the autocorrelation functions (ACF) for the second half of each chain. We see that ACF plots behave well and autocorrelation is quite small. We chose the initial points of each chain in order to speed up the convergence process of the MCMC, such that we could obtain it with less iterations. We can see in Table \ref{tab:posterior_with_Laplace} that the acceptance rates are in this case about of $23\%$. 

\begin{table}
\centering
\caption{Summary of the outcome divided by chain. $\tilde{\chi}^2$ taken as GPR. MH algorithm with half-Laplace prior. The unit for $m^I_g$, $ \langle m_g \rangle $ and the $97 \%$ quantile is $10^{-23} \; eVc^{-2}$. }
\begin{tabular}{| c || c | c | c | c | c | c | }
\hline
 Chain &  $m^I_g $ & $ \langle m_g \rangle $ & Acc. rate & $97.7 \% $  \\ [1.2ex]
\hline \hline 

1 & $0.108 $ &  $ 0.0156 $ & $23.4 \%$ &  $0.073 $  \\ \hline
2 & $0.071 $ &  $ 0.0153 $ & $23.5 \%$ &  $0.071 $  \\ \hline
3 & $0.093 $ &  $ 0.0156 $ & $23.2 \%$ &  $0.081 $  \\ \hline
4 & $0.040 $ &  $ 0.0156 $ & $23.8 \%$ &  $0.064 $  \\ \hline
5 & $0.023 $ &  $ 0.0160 $ & $24.3 \%$ &  $0.070 $  \\ \hline
\end{tabular}
\label{tab:posterior_with_Laplace}
\end{table}
\end{document}